# Design and Implementation of a ~25-Yr Pseudoprospective Earthquake Forecasting Experiment in China (AoyuX)

**Jiawei Li[1], Qingyuan Zhang[1,2,3,4], and Didier Sornette[1*]**

[1]Institute of Risk Analysis, Prediction and Management (Risks-X), Academy for Advanced Interdisciplinary Studies, Southern University of Science and Technology (SUSTech), Shenzhen, China.

[2]School of Systems Science, Beijing Normal University, Beijing, China.

[3]International Academic Center of Complex Systems, Beijing Normal University, Zhuhai, China.

[4]Department of Systems Science, Faculty of Arts and Sciences, Beijing Normal University, Zhuhai, China.

*Corresponding author: Didier Sornette (dsornette@ethz.ch)

## Declaration of Competing Interests

The authors acknowledge there are no conflicts of interest recorded.

## Abstract

Forecast models in statistical seismology are commonly evaluated with log-likelihood (*LL*) scores of the full distribution $P(n)$ of earthquake numbers, yet fat tails and out-of-range observations (realized counts exceeding all simulations) can bias model ranking. We develop a tail-aware evaluation framework that estimates cell-wise $P(n)$ using adaptive Gaussian kernel density estimation and tests three strategies for handling out-of-range counts (power law extrapolation, fixed minimum probability, and exclusion). Using the AoyuX platform, we perform a ~25-year (2000–2023) month-by-month pseudo-prospective forecast experiment in the China Seismic Experimental Site (CSES), comparing Epidemic-Type Aftershock Sequence (ETAS) model with a homogeneous background ($\mathrm{ETAS}_{\mu}$) to a spatially heterogeneous variant ($\mathrm{ETAS}_{\mu(x,y)}$) across six spatial resolutions and five magnitude thresholds; each forecast window is supported by 100,000 simulations. Empirical probability density functions (PDFs) of counts per cell are well described by power laws $P(n) \sim n^{-(1+a)}$ with exponents $a \approx 1.02 \pm 0.23$ across all settings. Using previous theoretical results, this provides a robust estimate of the productivity exponent, $\alpha = 0.83 \pm 0.19$ using a $b$-value equal to $0.85 \pm 0.002$, providing a valuable quantification of this key parameter in aftershock modeling. Model ranking is sensitive to how the tail of the full distribution $P(n)$ of earthquake counts is treated: power law extrapolation is both theoretically justified and empirically the most robust, while the apparent superiority of the fixed-probability approach in a few coarse low-magnitude cases can be attributed to noise. Cumulative information gain (CIG) shows that $\mathrm{ETAS}_{\mu(x,y)}$ outperforms $\mathrm{ETAS}_{\mu}$ in data-rich configurations (e.g., $M_c = 3.0$ to 3.5 and whole-region tests), with positive mean information gain spatially concentrated in regions where $M \geq 5$ earthquakes occurred during the study period, whereas in data-poor settings stochastic fluctuations dominate. A coefficient-of-variation analysis of per-window log-likelihood differences distinguishes genuine upward trends in CIG from noise-dominated fluctuations. By aligning a fat-tail-aware scoring methodology with an open testing platform, our work advances fair and statistically grounded assessment of earthquake forecasting models for the CSES and beyond.

## 1 Introduction

Advancing earthquake forecasting (probabilistic) and/or prediction (deterministic) benefits from a transparent and iterative research paradigm: build models that explain observed seismicity; test them prospectively or pseudo-prospectively against withheld or genuinely future data; use the resulting feedback to validate assumptions; and then retain, revise, or replace model components before repeating the cycle. Such a loop of construction, testing, correction, and refinement underpins cumulative progress in the field (Kagan & Jackson, 2000; Werner et al., 2011; Ogata, 2022). Crucial to this paradigm is rigorous, unbiased performance evaluation: determining, with statistical significance, whether a candidate model truly outperforms a designated benchmark (the null hypothesis, $H_0$) rather than relying on ad hoc or subjective criteria. The international Collaboratory for the Study of Earthquake Predictability (CSEP) and RichterX global earthquake forecasting platform formalized this ethos with standardized tests, scoring rules, and community benchmarks, demonstrating both the value and the challenges of fair model comparison (Schorlemmer et al., 2007; 2010; Taroni et al., 2018; Kamer et al., 2021; Nandan et al., 2021a; Bayliss et al., 2022).

In practice, a widely adopted evaluation framework is to run pseudo-prospective experiments over consecutive, non-overlapping time windows. Genuine prospective forecasting, evaluation on data not yet observed, should be stressed, with pseudo-prospective experiments serving as a preparatory or complementary step when fully prospective deployment is not yet feasible. For each time window and for each model-benchmark pair, one computes the full probability distribution or probability density function (PDF) $P(n)$ of earthquake numbers $n$ at the grid-cell level (given a chosen magnitude threshold $M_t$ and spatial resolution) and then scores the model by the log-likelihood ($LL$) obtained as the logarithm of the products over all cells of the $P(n_i)$'s, where $n_i$ are the observed counts in each $i$-th grid (see Section 2.2 for more details). Forecasting and scoring the distribution, rather than only its mean rate, has been shown to be both fair and robust, and it aligns naturally with aftershock-cascade physics (Ogata et al., 2013; Nandan et al., 2019a; 2019b). The approach is compatible with a range of models used operationally, including Epidemic-Type Aftershock Sequence (ETAS) model and its variants (Zhuang, 2011; Werner et al., 2011; Hainzl et al., 2024; Mizrahi et al., 2024; Han et al., 2025).

Three methodological issues arise when estimating the full PDF $P(n)$:

**(1) Randomness.** Because $P(n)$ is typically derived from a large but finite ensemble of simulated catalogs, sampling variability introduces artificial roughness into the distribution. To mitigate this effect, studies have traditionally adopted parametric count models, such as Poisson, negative-binomial or Beta-negative binomial approximations, or, more recently, non-parametric smoothers like adaptive kernel density estimators (Harte, 2017; Kagan, 2010; 2017; Molchan & Varini, 2024; Nandan et al., 2019a; Taroni et al., 2023; Werner et al., 2011). The use of the Poisson distribution is particularly natural when a forecast specifies only the mean rate, since, under the maximum entropy principle, the Poisson distribution is the most parsimonious choice: it imposes the fewest assumptions beyond the mean number of events. However, once a forecast conveys information beyond the mean rate, restricting the distribution to Poisson is no longer theoretically justified. In this context, non-parametric approaches, especially adaptive kernel density estimators, have emerged as effective tools for stabilizing likelihood-based evaluations while minimizing model-induced bias (Nandan et al., 2019a; 2021a; 2021b).

**(2) Fat-tailed (or power law-tailed) behavior.** The frequency distribution of counts across cells is empirically close to a power law $P(n) \sim n^{-(1+a)}$ in its upper tail, with an exponent $a$ often smaller than 2, qualifying the terminology "heavy-tailed" (Embrechts et al., 1997; Sornette, 2006). The term "fat-tailed" for distributions is broader and refers to distributions that have right-end tails that are thicker than that of a Gaussian and an exponential distribution. To distinguish with the term "heavy-tail," the term "fat-tailed distribution" may usually refer to a power law with exponent $a$ larger than 2, in which case the variance is finite, or it may have a non-power law tail such as a stretched exponential (Laherrère & Sornette, 1998). We shall use the term "fat-tailed" in the broader sense.

The power law tail of the PDF $P(n)$ reflects clustered triggering and cascade dynamics (Kagan, 2017; Nandan et al., 2019a; Nandan et al., 2021a; see also Figures 6 & 7 in the present study as examples). This power law property is in fact predicted by theory for ETAS models of earthquake triggering (Saichev et al., 2005; Saichev & Sornette, 2006; 2007). Yet most evaluation pipelines, especially those relying on non-parametric smoothing, do not explicitly constrain or extrapolate the tail (Nandan et al., 2019a; Nandan et al., 2021b; Mizrahi et al., 2024; Han et al., 2025).

**(3) Structural limitations of ETAS-based simulations.** In addition to the statistical fat-tailed behavior of earthquake-count distributions, out-of-range observations may also reflect a more fundamental limitation of standard ETAS formulations. In cells impacted by aftershocks of great earthquakes, observed counts can exceed the maximum of the simulated ensemble, requiring an out-of-range tail treatment (e.g., extrapolation vs. truncation). How such tail handling affects comparative model ranking has not been systematically quantified. Although ETAS models are effective in describing short-term triggering and aftershock cascades, they generally assume a background seismicity rate that is stationary in time and space. They do not explicitly incorporate long-term tectonic loading, fault-specific recurrence behavior, or time-dependent stress accumulation. As a result, large but physically plausible earthquakes that are absent from the calibration catalog may be assigned extremely low probability, or may be effectively missing from the simulated ensemble. This limitation is particularly relevant for fault systems such as the Longmenshan fault zone, where the 2008 Wenchuan $M_S$ 8.0 earthquake and its aftershock sequence strongly reshaped the seismicity pattern in the adjacent region. Therefore, the out-of-range problem should not be interpreted solely as a numerical artifact of finite simulations or as a consequence of fat-tailed count distributions. It also highlights the structural incompleteness of purely ETAS-based forecasting models when rare, fault-controlled large earthquakes are not represented in the calibration history.

Increasing the number of simulated catalogs can, to some extent, mitigate the three issues mentioned above. However, simply enlarging the ensemble size is not a practical solution in real applications. Doing so would not only lead to a substantial waste of computational resources but also significantly prolong the running time, thereby raising the overall computational cost. More importantly, even an arbitrarily large number of simulated catalogs cannot fully guarantee a perfect resolution of the underlying problems, as the fundamental limitations are methodological rather than merely statistical.

This study addresses that gap. We develop an evaluation framework that (i) estimates the cell-wise probability distribution $P(n)$ of earthquake numbers $n$ with adaptive smoothing, (ii) explicitly incorporates fat-tail behaviour with principled extrapolation for out-of-range observations, and (iii) tests how sensitive model ranking are to these choices. Our analysis complements recent advances in short-term forecasting across diverse regions and model classes, including ETAS generalizations and alternative point-process formulations (Zhuang, 2011; Werner et al., 2011; Davis et al., 2024; Hainzl et al., 2024; Mizrahi et al., 2024; Han et al., 2025).

To enable transparent and repeatable testing at scale, we introduce AoyuX, an online research platform dedicated to earthquake predictability studies in China. Developed by the Institute of Risk Analysis, Prediction and Management (Risks-X) at the Southern University of Science and Technology (SUSTech) with the China Earthquake Networks Center and other institutions as partner organizations, AoyuX integrates state-of-the-art statistical seismology, centered on ETAS-type models that jointly forecast time-space-magnitude probabilities, with an extensible architecture for benchmarking, visualization, and (future) model-comparison challenges. The initial operational focus is the China Seismic Experimental Site (CSES) in Sichuan-Yunnan region, one of the nation's most instrumented and seismically active regions (Li et al., 2022; Wu et al., 2019; Wu & Li, 2021a; 2021b; Wu, 2022). Since June 2024, AoyuX has issued monthly, quarterly, semi-annual, and annual forward forecast bulletins for the CSES (see Data and Resources), thereby establishing a sustained prospective and pseudo-prospective testing environment. Looking ahead, AoyuX will fuse geophysical observations with AI-assisted and physics-based components (e.g., Zhan et al., 2024; Zhang et al., 2025) to support mid-term, short-term, and imminent forecasting, and to facilitate communication between experts and the public.

In the present study, we conducted a ~25-year month-by-month pseudo-prospective forecasting experiment for the China Seismic Experimental Site (CSES), using the ETAS model together with a spatially heterogeneous variant in which the background seismicity rate varies across space. We further evaluated the impact of innovative fat-tail treatments for out-of-range observations and investigated under which conditions the cumulative information gain (CIG) curves exhibit a clear long-term upward trend, or alternatively, when such a trend is obscured by stochastic fluctuations. We also examined the spatial distribution of mean information gain (MIG) to characterize where the ETAS model with spatially heterogeneous background seismicity differs from its homogeneous background counterpart. By integrating a fat-tail-aware scoring methodology with a dedicated open testing platform, our work advances both the methodological foundations and the research infrastructure required for fair and statistically rigorous assessment of earthquake forecasting models in China and beyond.

## 2 Methods

### 2.1 The Epidemic-Type Aftershock Sequence (ETAS) model for earthquake forecasting

The Epidemic-Type Aftershock Sequence (ETAS) model is a widely used formulation of the self-exciting Hawkes point process, specifically adapted to characterize seismicity. Unlike Markovian processes, the ETAS model generally exhibits non-Markovian behavior when its temporal memory kernel deviates from an exponential form (Hawkes & Oakes, 1974; Kagan & Knopoff, 1981, 1987; Ogata, 1988, 1998; Kanazawa & Sornette, 2020; 2024). At any spatio-temporal location ($x$, $y$, $t$), the model defines the expected seismicity rate, or intensity function $\lambda(x, y, t)$, which depends on the complete history of seismic events up to time $t$, denoted as $H_t$:

$$\begin{aligned}\lambda(x, y, t \mid H_t) &= \mu + \sum_{i:t_i<t} F(m_i) T(t-t_i) S(x-x_i, y-y_i, m_i) \\ &= \mu + \sum_{i:t_i<t} K e^{\alpha(m_i - M_{\mathrm{co}})} \frac{T_{\mathrm{norm}} e^{-\frac{t-t_i}{\tau}}}{(t-t_i+c)^{1+\omega}} \frac{S_{\mathrm{norm}}}{[(x-x_i)^2+(y-y_i)^2+d e^{\gamma(m_i-M_{\mathrm{co}})}]^{1+\rho}}\end{aligned} \tag{1}$$

Here, the intensity $\lambda$ consists of a background rate $\mu$ and contributions from all prior earthquakes. Each past event $i$, characterized by its magnitude $m_i$ and location ($x_i$, $y_i$, $t_i$), influences future seismicity through a separable spatio-temporal triggering kernel, weighted by a fertility (or productivity) function $F(m_i)$, parameterized by constants $K$ and $\alpha$. The spatio-temporal triggering kernel is typically defined as the product of two components: a temporal kernel described by the Omori-Utsu law $T(t - t_i)$, with parameters $c$, $\omega$, and $\tau$; and a spatial kernel represented by a Green's function with parameters $d$, $\gamma$, and $\rho$, together describing the temporal and spatial dependencies of aftershock triggering. The normalization constants $T_{\text{norm}}$ and $S_{\text{norm}}$ ensure that the temporal and spatial kernels are valid PDFs. This formulation reflects a commonly adopted choice for the spatio-temporal triggering kernel in ETAS models (Nandan et al., 2021b; 2022), although alternative kernel forms have also been proposed in the literature (Helmstetter & Sornette, 2002; Felzer et al., 2002, 2003; Zhuang et al., 2002, 2005).

The ETAS model admits two conceptually distinct yet mathematically equivalent interpretations (Helmstetter & Sornette, 2002; Li et al., 2025a):

**(a) Branching interpretation** (Hawkes & Oakes, 1974; Kagan, 1991; Sornette & Werner, 2005b). Each observed earthquake can be probabilistically linked to a parent event, forming a cascade of primary and secondary aftershocks in a tree-like structure. In this framework, background events follow a Poisson process, while each event may trigger its own sequence of aftershocks. Every tree configuration is associated with a probability that quantifies the likelihood of a specific parent-offspring association, given the observed catalog. These probabilities are derived from the ETAS intensity function $\lambda(x, y, t)$ (Zhuang et al., 2002).

**(b) Collective influence interpretation** (Helmstetter & Sornette, 2002). Each non-background event arises from the cumulative triggering effect of all preceding earthquakes. In this view, the intensity function $\lambda(x, y, t)$ represents the combined influence of historical seismicity at a given location and time, superimposed on the background rate.

The mathematical equivalence between these two interpretations was rigorously established by Hawkes & Oakes (1974). Sornette & Werner (2005b) further offered an intuitive explanation for this equivalence, based on the combination of (i) the formulation of the ETAS seismicity rate expressed as a linear superposition of contributions from prior events and of (ii) the exponential form of the point process hazard rate. This dual perspective enables the ETAS model to function both as a generative engine of synthetic earthquake catalogs and as a predictive tool for time-dependent seismic hazard assessment.

In the present study, we employ two variants of the ETAS model that share an identical triggering kernel, with the sole difference being the treatment of the background seismicity rate (Nandan et al., 2021b; 2022):

(1) ETAS model with a constant background rate ($\text{ETAS}_{\mu}$). This variant, defined by Equation (1), assumes a spatially homogeneous background rate $\mu$. We adopt it as the benchmark model in the present study. Its estimation involves determining the following set of model parameters: $\{\mu, K, \alpha, c, \omega, \tau, d, \gamma, \rho\}$.

(2) ETAS model with spatially variable background rate ($\text{ETAS}_{\mu(x,y)}$). In this augmented variant, the background rate $\mu(x, y)$ is treated as a nonparametric function that captures spatial heterogeneity of the background events. Its estimation follows a weighted kernel approach, as introduced by Nandan et al. (2021b, 2022):

$$\mu(x,y) = \frac{1}{T}\sum_{i=1}^{N_{\text{eq}}} IP_i \cdot \frac{QD^{2Q}}{\pi}\frac{1}{[(x-x_i)^2+(y-y_i)^2+D^2]^{1+Q}} \tag{2}$$

where the summation is over all $N_{eq}$ earthquakes in the catalog, weighted by their probability $IP_i$ of being background events (Zhuang et al., 2002). The normalization by the catalog duration $T$ ensures that $\mu(x, y)$ represents the seismicity rate per unit time, while the factor $QD^{2Q}/\pi$ normalizes the kernel per unit area at location $(x, y)$. The choice of a power law kernel is motivated by its demonstrated superiority over the commonly used Gaussian kernel in previous studies (e.g., Helmstetter et al., 2007; Nandan et al., 2021a). Its estimation involves determining the following set of parameters: $\{Q, D, K, \alpha, c, \omega, \tau, d, \gamma, \rho\}$.

The parameter calibration of both variants is carried out using the extended Expectation-Maximization (EM) algorithm (Veen & Schoenberg, 2008). We refer readers to Nandan et al. (2021b, 2022) for full methodological details.

2.2 Pseudo-prospective forecasting experiments

The pseudo-prospective forecasting experiment is conducted over the testing period from January 1, 2000, to December 31, 2023. Each forecast window spans one calendar month, and the windows are non-overlapping. Specifically, for a given forecast window (e.g., a particular month in a given year), the earthquake catalog prior to that month is used to calibrate the model parameters. Based on the calibrated parameters, 100,000 synthetic earthquake catalogs are simulated for the forecast window. These synthetic catalogs are used to estimate the occurrence probability of the future earthquakes. The process is then repeated for each subsequent month in the testing period. To assess the spatial performance of the models, the study region is discretized using a geodesic triangular tessellation on the Earth's surface accounting for its spherical geometry, with six different spatial resolutions defined by equal-area triangles of approximately 195 $km^2$, 779 $km^2$, 3,117 $km^2$, 12,470 $km^2$, 49,883 $km^2$, and the entire CSES region as a single unit (approximately 780,000 km$^2$). In addition, five different magnitude thresholds $M_t$ are considered: 3.0, 3.5, 4.0, 4.5, and 5.0.

For each forecast window, magnitude threshold, and spatial resolution (space-time-magnitude bin), we compute the distribution of the number of simulated earthquakes in each triangular cell. The probability distribution $P(n)$, where $n$ denotes the number of earthquakes, is then estimated using the kernel-based smoothing method described in the next section. Accordingly, the probability of observing $n_i$ earthquakes in the $i$-th triangle is given by $P(n_i)$. Given a particular spatial resolution and magnitude threshold, the log-likelihood ($LL$) score of a model for a specific forecast window is defined as:

$$LL = \sum_{i=1}^{N} \log P(n_i) \tag{3}$$

where $N$ is the number of spatial cells at the specified resolution. The information gain (IG) is defined as the difference in log-likelihood ($LL$) scores between $\text{ETAS}_{\mu(x,y)}$ and $\text{ETAS}_{\mu}$. A higher IG indicates superior performance of the spatially heterogeneous $\text{ETAS}_{\mu(x,y)}$ model relative to the benchmark $\text{ETAS}_{\mu}$ model, and vice versa.

## 2.3 Distribution of forecasted earthquake numbers

To obtain a smooth distribution of forecasted earthquake numbers and minimize uncertainties arising from the finite number of simulations, we adopt an adaptive Gaussian kernel density estimation (KDE) method to represent the distribution of simulated earthquake numbers within each grid cell, given a specific forecast window, magnitude threshold $M_t$, and spatial resolution. The adaptive kernel density estimation method is a nonparametric approach that adjusts to varying data density by allowing the bandwidth to vary locally (Cranmer, 2001). Its key characteristic is that regions with higher data density are assigned narrower kernels, while sparser regions use wider kernels (Figure 1a), thereby improving estimation accuracy across the full distribution.

Although earthquake counts are discrete non-negative integers, we use the adaptive Gaussian KDE here as a smoothing device for the empirical distribution obtained from a large but finite ensemble of simulated catalogs, rather than as an assumption that the underlying count process is continuous. In practice, the KDE-smoothed density is evaluated only at the observed integer count $n_i$, and thus serves as a regularized estimate of the probability assigned to that observed count. This treatment is particularly useful when the number of synthetic catalogs is finite, because the raw empirical histogram can contain artificial gaps or strong sample-to-sample fluctuations (finite-simulation noise), especially in the upper tail where the number of simulated realizations is small. The continuous Gaussian kernel therefore provides a practical way to stabilize the likelihood calculation while retaining the nonparametric character of the simulated forecast distribution.

Discrete alternatives, such as discrete KDE or parametric count models including Poisson, negative-binomial, and beta-negative-binomial distributions, are possible in principle. However, these approaches would introduce additional assumptions about the functional form of $P(n)$. In particular, a Poisson model is appropriate mainly when the forecast specifies only the mean rate, whereas the simulated ETAS catalogs contain information about the full distribution of earthquake counts, including clustering-induced overdispersion and fat-tailed behavior. Negative-binomial-type models can represent overdispersion, but they may still be too restrictive for the irregular and model-specific distributions produced by aftershock cascades. We therefore adopt adaptive Gaussian KDE as a minimally parametric compromise that smooths the simulated distribution without forcing it into a prescribed count-family form. In our tests, using alternative smoothing methods do not materially change the results.

Smoothing can in principle affect likelihood estimation and hence model ranking. If the distribution is over-smoothed, sharp but meaningful differences between models may be attenuated; if it is under-smoothed, finite-simulation noise may be amplified and spuriously influence the log-likelihood. In the present study, this effect is mitigated in three ways. First, the same adaptive KDE procedure is applied consistently to both $\mathrm{ETAS}_{\mu}$ and $\mathrm{ETAS}_{\mu(x,y)}$ under each time-space-magnitude configuration. Second, the adaptive bandwidth reduces noise in sparse regions while preserving narrower kernels in high-density regions. Third, cases in which the observed count lies outside the simulated range are handled separately through the three tail-treatment strategies described below. Thus, the KDE mainly regularizes within-range probability estimates, whereas the sensitivity of model ranking to extreme tail treatment is explicitly evaluated in Sections 4.2 and 4.3.

In the present study, the kernel density $f(n)$ for the number of simulated earthquakes $n$ is given by:

$$f(n) = \frac{1}{N_{\mathrm{sim}}} \sum_{j=1}^{N_{\mathrm{sim}}} \frac{1}{h_j} K\left(\frac{n - n_j}{h_j}\right) \tag{4}$$

where $N_{\mathrm{sim}}$ is the number of synthetic catalogs (set to 100,000 in the present study), $n_j$ is the number of earthquakes in the $j$-th simulated catalog, $K(\cdot)$ is the standard Gaussian kernel with zero mean and unit variance, $h_j$ is the locally adaptive bandwidth for the $j$-th sample. To compute the pilot kernel density $f_0(n)$, each $h_j$ is initially set to a fixed bandwidth $h_0$, which is selected to minimize the mean integrated squared error (MISE) following a rule-of-thumb bandwidth selection (Scott, 2015). The adaptive bandwidths $h_j$ are then adjusted according to:

$$h_j = \frac{h_0}{f_0(n_j)} \tag{5}$$

where $f_0(n_j)$ is the pilot density estimate at $n_j$. This formulation ensures narrower bandwidths in regions where simulated earthquake counts are concentrated and broader bandwidths where data are sparse. In effect, each simulation contributes a Gaussian kernel in probability space, with its width adaptively scaled by the local data density, i.e., the number of synthetic catalogs yielding similar outcomes.

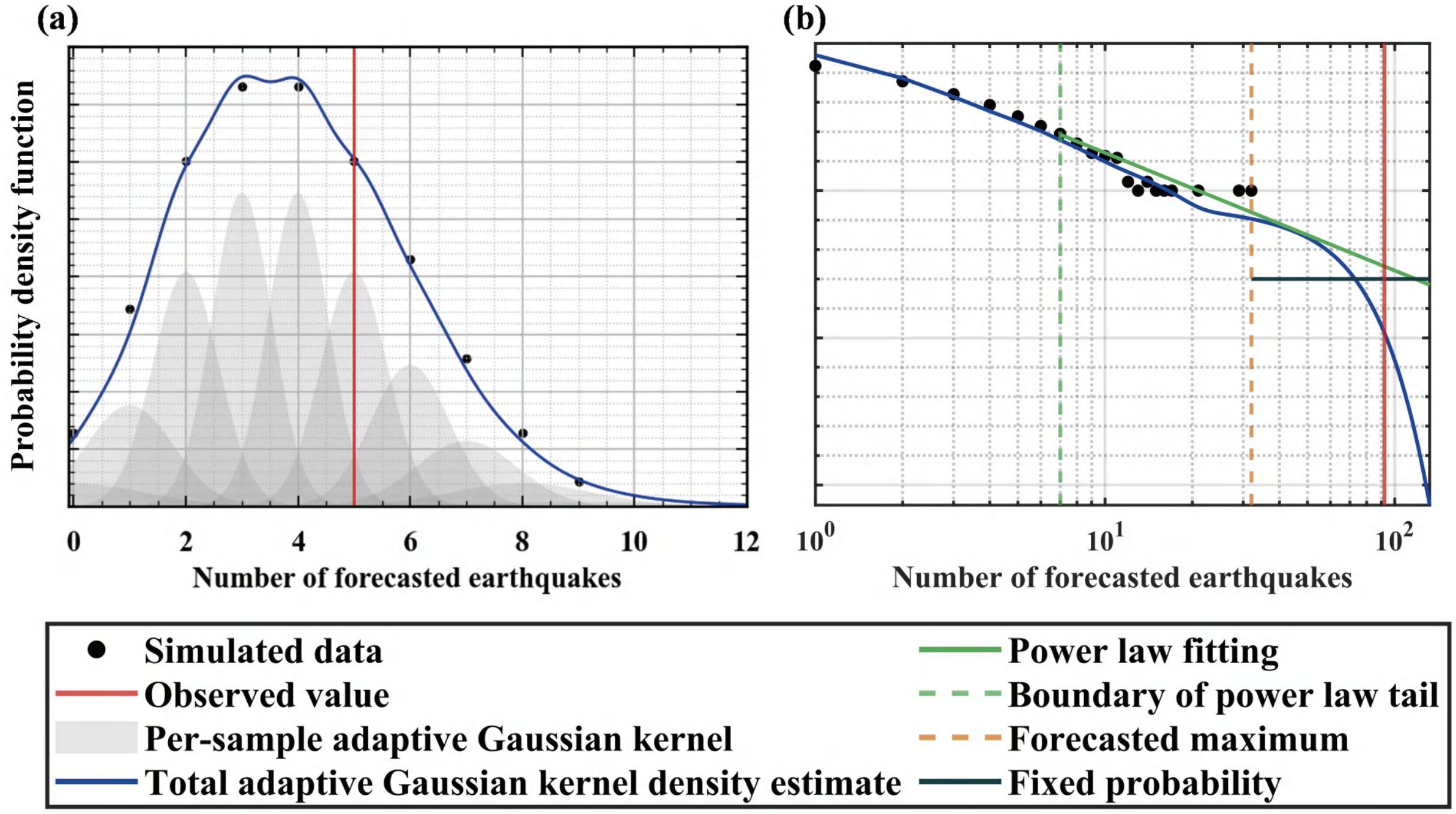

**Figure 1**. Illustration of the probability distribution for the number of forecasted earthquakes, shown for two scenarios: (a) where the observed value (red solid line) falls within the range of simulated values, and (b) where the observed value exceeds the maximum of the simulated distribution. The forecasted probability corresponding to the observed value $n_i$ in the $i$-th triangle is denoted as $P(n_i)$. In the present study, simulated data (black solid circles) are generated from $N_{\text{sim}}$ sets of synthetic catalogs ($N_{\text{sim}}$ = 100,000). The per-sample adaptive Gaussian kernels (gray shaded areas), each with its own bandwidth, are summed to construct the total adaptive Gaussian kernel density estimate (blue solid line), which is used to present the overall probability

distribution. When $n_i$ (red solid line) exceeds the forecasted maximum (orange dashed line) in panel (b), we propose three alternative strategies for assigning $P(n_i)$: (1) fitting a power law function (green solid line) to the upper 30% of the simulated tail data (to the right of the green dashed line); (2) assigning a fixed probability value (black horizontal line), equal to $1/(N_{\text{sim}})^2$; or (3) excluding $P(n_i)$ altogether in such cases.

**Alt-Text** Two-panel schematic illustrating how the probability of the observed earthquake count is evaluated relative to the simulated forecast distribution. In panel (a), the observed number of earthquakes (vertical red line) lies within the range of values generated from 100,000 synthetic catalogs (black dots). The overall forecast probability density (blue curve) is constructed by summing per-sample adaptive Gaussian kernels (gray shaded curves), each with its own bandwidth. The probability assigned to the observed count in the *i*-th spatial cell is denoted as $P(n_i)$. In panel (b), the observed count (red line) exceeds the maximum simulated value (orange dashed line). Three alternative approaches for assigning $P(n_i)$ in this case are illustrated: (1) extrapolating the upper 30% of the simulated tail with a fitted power law function (green curve, fitted to data right of the green dashed line); (2) assigning a fixed small probability equal to $1/(N_{\text{sim}})^2$ (black horizontal line); or (3) excluding the probability term when the observation falls beyond the simulated range.

However, when the observed number of earthquakes within a grid cell—given a specific forecast window, magnitude threshold $M_t$, and spatial resolution—exceeds the maximum value obtained from the simulated catalogs, the adaptive kernel method may yield serious biased estimates of the probability $P(n)$ (see, for example, Figures 1 and 5 in the present study and Figure 1 in Nandan et al. (2019a)). To address this issue, we propose three alternative strategies for assigning $P(n)$ in such cases (Figure 1b):

**(1) Power law extrapolation**. Fit a power law function to the upper 30% of the simulated tail data and use the extrapolated value as $P(n)$;

**(2) Fixed minimum probability**. Assign a fixed probability value equal to $1/(N_{\text{sim}})^2$;

**(3) Exclusion**. Exclude the corresponding $P(n)$ value entirely from the likelihood

computation.

In the following sections, we evaluate the impact of these three strategies on model performance in forecasting experiments.

## 3 Seismicity in China Seismic Experimental Site (CSES)

Situated in a seismically active region (Li et al., 2020; 2021; 2022), the China Seismic Experimental Site (CSES) is designed to integrate dense, multidisciplinary observations to investigate the complex physical processes underlying earthquake generation (Wu et al., 2019; Wu & Li, 2021a; 2021b; Wu, 2022). By combining geological, geophysical, and geochemical data, the site enables a comprehensive investigation of seismic hazards. This integrative approach facilitates the development and validation of predictive models, thereby contributing significantly to earthquake forecasting efforts in China. The CSES region has long maintained a dense and high-quality seismic monitoring network, resulting in an extended record of reliable earthquake catalogs (Li et al., 2023; 2025c). As shown in Figure 2, the spatio-temporal distribution of completeness magnitude $M_c$ indicates that the CSES catalog has been nearly complete for events with $M \geq 3$ since 1985. Short-lived degradations of completeness are observed mainly around several strong earthquake sequences, especially around 1988, 1995, and following the 2008 Wenchuan $M_S$ 8.0 earthquake, when the completeness probability for $M \geq 3$ dropped to approximately 50% to 97.8% (Figure 2c). These transient episodes are typical of short-term aftershock incompleteness caused by extremely high aftershock rates and overlapping seismic signals immediately after large earthquakes. Therefore, although the catalog is not perfectly complete at all times, the use of 1985 as the starting time for ETAS calibration does not introduce a persistent completeness bias at the adopted catalog cut-off magnitude at $M_{co} = 3.0$.

Instead, it provides a longer historical record for constraining the background seismicity rate and the triggering history prior to the testing period.

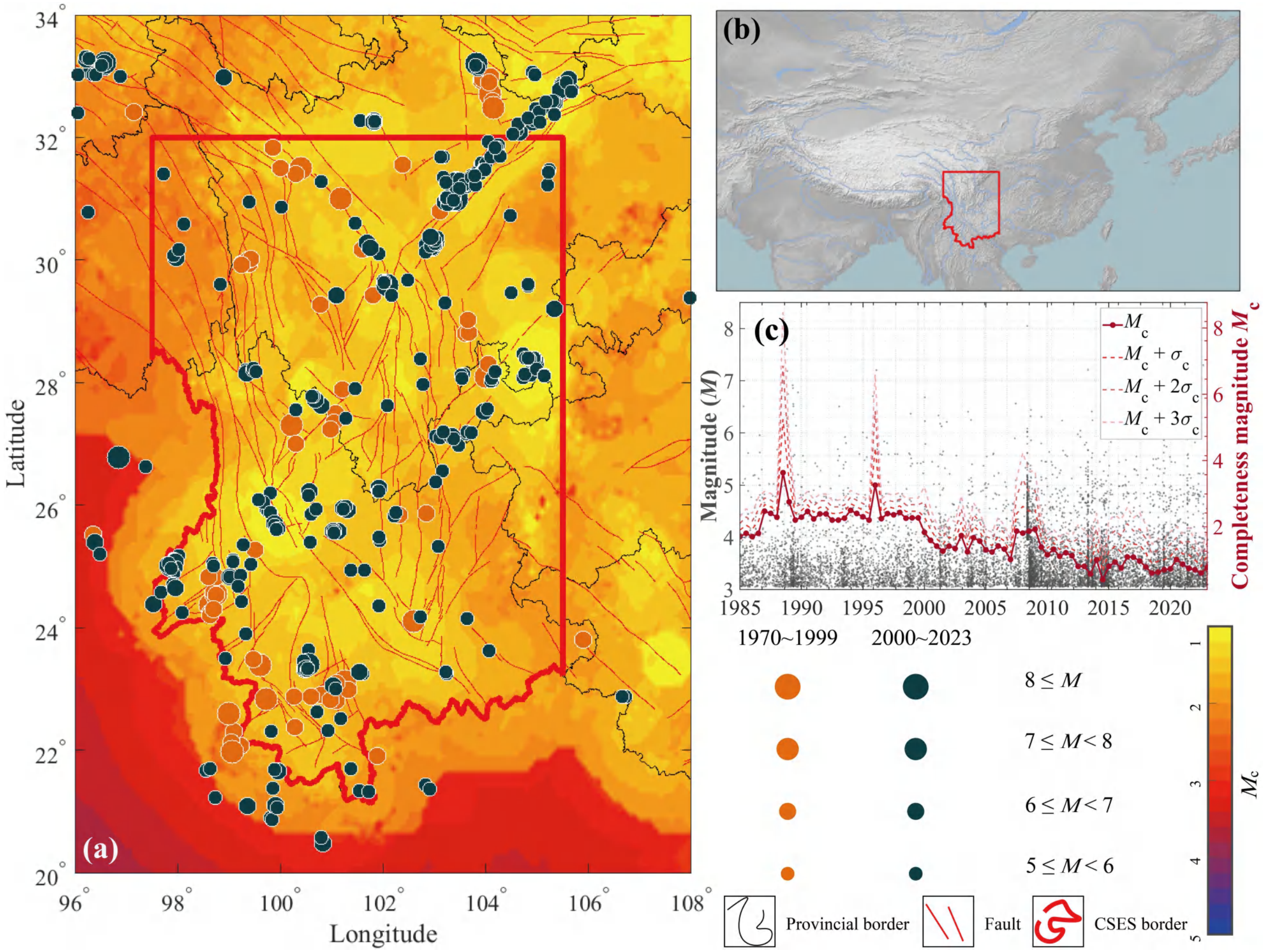

**Figure 2**. Spatio-temporal distribution of seismicity and completeness magnitude ($M_c$) in the China Seismic Experimental Site (CSES), located in the Sichuan-Yunnan region. (a) Spatial distribution of earthquakes from 1970 to 2023, completeness magnitude ($M_c$) since 2009, and active faults within the CSES. The background map of $M_c$ is adapted from Li et al. (2023). Fault locations are taken from Deng et al. (2003). (b) Location of the CSES shown on a broader regional map. (c) Temporal evolution of $M_c$ (right $y$-axis) and the magnitude-time (MT) plot of earthquakes with $M \geq 3$ (left $y$-axis) since 1985. The estimation of $M_c$ was performed using a half-year step and one-year window. Red dashed lines with decreasing color indicate completeness probabilities of 84.2% ($M_c + \sigma_c$), 97.8% ($M_c + 2\sigma_c$), and 99.9% ($M_c + 3\sigma_c$), respectively. $M_c$ and $\sigma_c$ are estimated by the method proposed by Wang et al. (2025) and Li et al.,

(2026). The earthquake magnitudes shown in the magnitude-time plot have been perturbed by a uniformly distributed random error within the range [-0.05, 0.05].

**Alt-Text** Three-panel figure showing the spatial and temporal characteristics of seismicity and completeness magnitude ($M_c$) in the China Seismic Experimental Site (CSES) in the Sichuan-Yunnan region. Panel (a) presents the spatial distribution of earthquakes from 1970 to 2023, overlaid on a background map of $M_c$ since 2009 and mapped active faults. Panel (b) shows the geographic location of the CSES within a broader regional context. Panel (c) displays the temporal evolution of $M_c$ (right vertical axis) together with a magnitude-time plot of earthquakes with $M \geq 3$ since 1985 (left vertical axis). $M_c$ is estimated using a one-year moving window with a half-year step. Red dashed curves with progressively lighter shading indicate completeness probability levels of 84.2% ($M_c + \sigma_c$), 97.8% ($M_c + 2\sigma_c$), and 99.9% ($M_c + 3\sigma_c$). Earthquake magnitudes in the magnitude-time plot are perturbed by a uniform random error within the range [-0.05, 0.05] to reduce visual overlap.

Nevertheless, short-term incompleteness after strong earthquakes can affect ETAS parameter estimation and should be acknowledged. Missing early aftershocks may reduce the apparent number of triggered events, thereby biasing the estimated triggering productivity, branching ratio, and early-time Omori-Utsu decay parameters. In particular, the productivity-related parameters $K$ and $\alpha$, as well as the temporal kernel parameters $c$ and $\omega$, can be affected because the earliest and densest part of an aftershock sequence is under-sampled. At the same time, missing events can break parent-offspring links in the inferred triggering structure, causing some events that are physically triggered to be classified as independent background events. This may lead to an apparent increase in the estimated background component, especially for the spatially variable background rate $\mu(x, y)$. Such effects have been discussed extensively in ETAS-based studies of incomplete catalogs and short-term aftershock incompleteness (Sornette & Werner, 2005b; Omi et al., 2014; Zhuang et al., 2017; Mizrahi et al., 2021; Hainzl, 2022; Hainzl et al., 2024). In the present study, these effects are expected to be temporally localized around a few major earthquake sequences and may partly explain the abrupt parameter changes observed in Figure 3, but they do not undermine the overall use of $M_{co} = 3.0$ for the long-term calibration catalog.

The earthquake catalogs used in the present study were provided by the China Earthquake Networks Center (CENC). The catalog of earthquakes with $M \geq 3.0$ in the CSES region during 1970-2023 is available as Data Set S1 in Li et al. (2025a). The CSES catalog includes a mixture of magnitude types, primarily local magnitude ($M_L$) and surface-wave magnitude ($M_S$), which introduces potential complications due to magnitude-type heterogeneity. To evaluate the possible influence of this issue, we performed a simple sensitivity test by varying the relative contributions of different magnitude types in a pseudo-prospective forecasting experiment. The results indicate that magnitude-type heterogeneity may introduce some variability, but it does not systematically alter the overall CIG patterns or dominate the model-comparison results. The magnitude bin size is set to 0.1. Temporally, the catalog is divided into two segments: (1) an auxiliary catalog from January 1, 1970, to December 31, 1984, used to avoid misclassifying early events in the primary catalog as background events in the ETAS model; and (2) the primary catalog, extending from January 1, 1985, to the day before the forecast window, used for calibrating the ETAS model parameters.

## 4 Results

In this section, we present two sets of results: (1) the calibrated parameters of the ETAS models, and (2) the outcomes of pseudo-prospective earthquake forecasting experiments based on the calibrated models.

### 4.1 Calibrated parameters of the ETAS model

To examine the temporal stability and evolution of ETAS model parameters, we calibrate both $\mathrm{ETAS}_{\mu(x,y)}$ and the homogeneous counterpart $\mathrm{ETAS}_{\mu}$ over a series of monthly forecast windows. For each calibration step, the primary catalog includes all events from January 1, 1985, up to the day preceding the forecast window. In addition, a 15-year auxiliary catalog covering the period from January 1, 1970, to December 31, 1984, is appended to provide historical context and reduce the likelihood of misclassifying earlier events as background earthquakes.

Figure 3 shows the temporal evolution of the estimated parameters for both $\mathrm{ETAS}_{\mu(x,y)}$ and $\mathrm{ETAS}_{\mu}$. Overall, the two models yield noticeably different parameter estimates, with $\mathrm{ETAS}_{\mu(x,y)}$ identifying significantly more background events than $\mathrm{ETAS}_{\mu}$, primarily as a result of allowing for a spatially heterogeneous background seismicity, which is otherwise likely to be misclassified as triggered seismicity in $\mathrm{ETAS}_{\mu}$. Furthermore, from 2000 to 2023, the differences between the two models in several key parameters remain relatively stable over time. Specifically, the fertility law parameters ($\log_{10}K$ and $\alpha$), the spatial kernel parameter ($\log_{10}d$), and the branching ratio $h$ consistently exhibit similar difference amplitudes between $\mathrm{ETAS}_{\mu(x,y)}$ and $\mathrm{ETAS}_{\mu}$. In contrast, the gap between the two models in terms of the cumulative number of background events and $\log_{10}\tau$ gradually widens over time. On the other hand, the differences in $\rho$, $\gamma$, $\log_{10}c$, and $1+\omega$ tend to decrease with time and, in some cases, nearly vanish. Notably, abrupt changes in several parameter time series coincide with the occurrence of major earthquakes in the region. Examples include the 2008 Wenchuan $M_S$ 8.0 earthquake (May 12), the 2013 Lushan $M_S$ 7.0 earthquake (April 20), and the 2017 Jiuzhaigou $M_S$ 7.0 earthquake (August 8), among others.

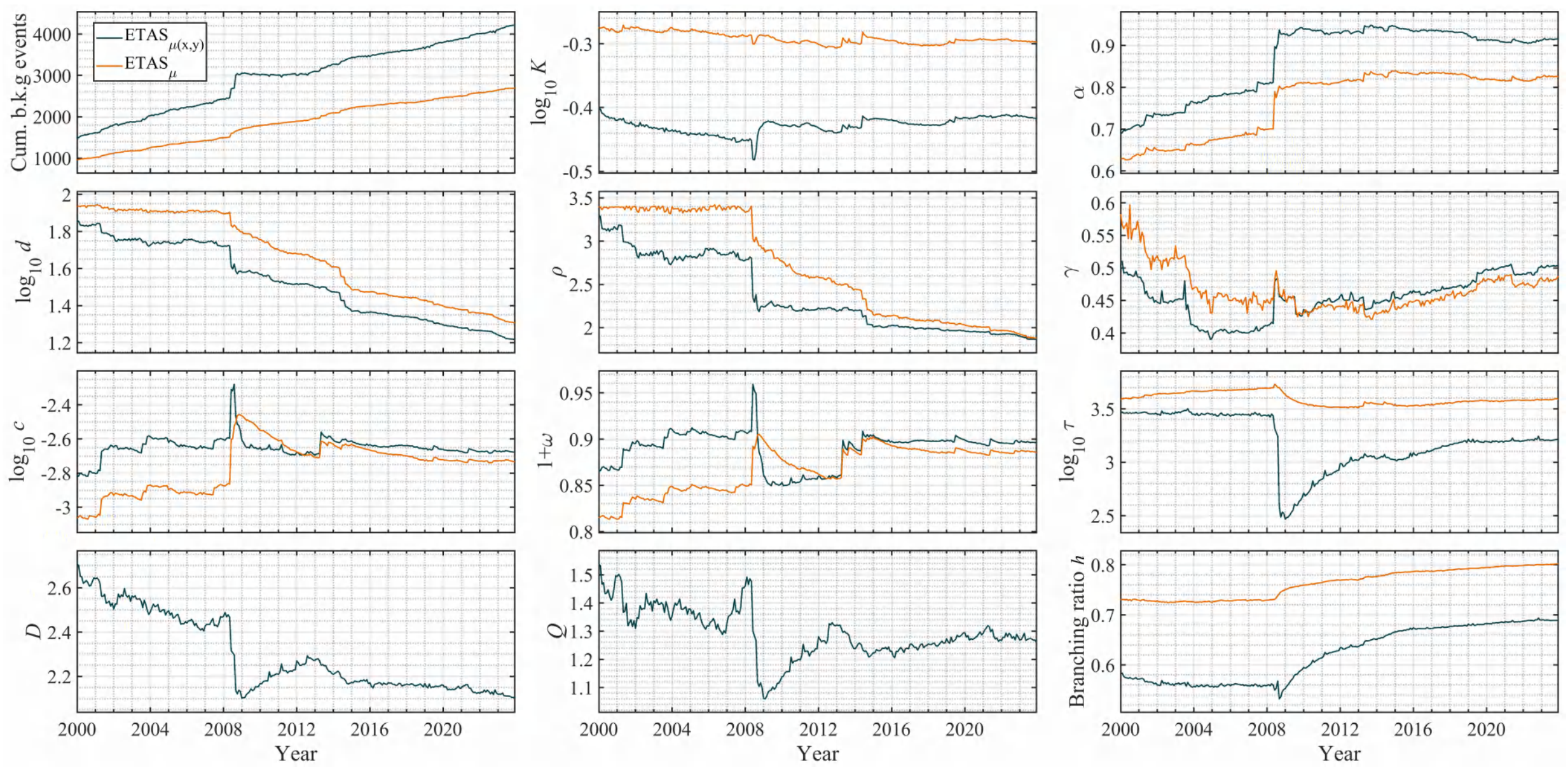

**Figure 3**. Temporal evolution of the calibrated parameters in the ETAS model with spatially variable background seismicity ($ETAS_{\mu(x,y)}$) and in the ETAS model with a constant background rate ($ETAS_{\mu}$). For model calibration, the primary catalog spans from January 1, 1985, to the day before each monthly forecast window, and is preceded by a 15-year auxiliary catalog (from January 1, 1970, to December 31, 1984) to prevent misclassification of early events in the primary catalog as background events. Note that the parameters $D$ and $Q$ are specific to $ETAS_{\mu(x,y)}$ and are not included in $ETAS_{\mu}$.

**Alt-Text** Time series comparing the monthly calibrated parameter values of two ETAS model configurations: one with spatially variable background seismicity, $ETAS_{\mu(x,y)}$, and one with a spatially constant background rate, $ETAS_{\mu}$. For each monthly forecast, parameters are estimated using a primary catalog from 1 January 1985 up to the day preceding the forecast window, supplemented by a 15-year auxiliary catalog (1 January 1970 to 31 December 1984) to reduce misclassification of early primary events as background events. The figure shows the temporal evolution of shared ETAS parameters in both models, while the spatial kernel parameters $D$ and $Q$ appear only for $ETAS_{\mu(x,y)}$ and are not defined for $ETAS_{\mu}$.

### 4.2 Cumulative information gain (CIG) of $ETAS_{\mu(x,y)}$ over $ETAS_{\mu}$

The calibrated parameters prepared for each forecast window were used to generate synthetic earthquake catalogs based on the first of the two conceptually distinct yet mathematically equivalent interpretations of the ETAS model, that is, the branching interpretation (see the second paragraph of Section 2.1). Specifically, for each forecast window, the ETAS parameters calibrated using the catalog prior to that window were used to simulate the seismicity expected during the target month. The simulated seismicity in each forecast window consists of three components: earthquakes triggered by events that occurred before the beginning of the forecast window, background earthquakes generated during the forecast window, and earthquakes triggered by these background events within the same window. In this way, the triggering memory inherited from the pre-window catalog is explicitly propagated into the forecast period, while newly generated background events are also allowed to initiate their own aftershock sequences. To reduce boundary effects in both model calibration and simulation, we used an auxiliary region extending 100 km beyond the CSES boundary; however, only earthquakes occurring within the CSES region were retained for forecast evaluation. The *b*-value used for magnitude sampling was treated in the same way as the other ETAS parameters: for each forecast window, it was estimated from the catalog prior to that window and then used in the corresponding simulations. For each forecast window, 100,000 synthetic catalogs were generated in this study. All simulated catalogs produced by the two ETAS models are openly available through Li et al. (2025d).

Figure 4 presents the time series of cumulative information gain (CIG) of $\mathrm{ETAS}_{\mu(x,\,y)}$ over $\mathrm{ETAS}_{\mu}$ from 2000 to 2023. Results are shown for six spatial resolutions, five magnitude thresholds ($M_t$), and three strategies for assigning $P(n_i)$ when the observed value $n_i$ in the $i$-th triangle exceeds the maximum value obtained from the simulations. The green, black, and orange curves correspond to strategies (1), (2), and (3) described in Section 2.3, respectively. We also provide, in the Data and Resources section, adaptive Gaussian kernel density fits for both models across the five magnitude thresholds for the full CSES region.

Table 1 reports the percentage of out-of-range grid cells across all 288 forecast windows for each combination of spatial resolution and magnitude threshold $M_t$. For the gridded spatial resolutions, the percentages are generally low at $M_t = 3$ and $M_t = 3.5$, but these configurations contain a large number of spatial cells, so the absolute number of out-of-range cases is still non-negligible. The percentage increases markedly at higher magnitude thresholds, especially for $M_t = 4.5$ and $M_t = 5$, where the forecasted count distributions become increasingly sparse. These results show that out-of-range observations are not restricted to a single extreme event, but occur across different space-magnitude configurations, particularly in sparse-count settings. This motivates the comparison of the three tail-treatment strategies in Figure 4.

**Table 1.** Percentage of out-of-range grid cells across all 288 time windows for each combination of spatial resolution and magnitude threshold $M_t$. Here, an out-of-range grid cell refers to a case in which the maximum earthquake count obtained from the simulated catalogs is smaller than the corresponding observed earthquake count. The percentage is calculated relative to the total number of grid cells in each configuration.

| Spatial resolution | $M_t = 3$ | $M_t = 3.5$ | $M_t = 4$ | $M_t = 4.5$ | $M_t = 5$ |
|---|---|---|---|---|---|
| 195 km$^2$ | 0.36% | 0.51% | 0.73% | 1.36% | 1.74% |
| 779 km$^2$ | 0.15% | 0.22% | 0.40% | 0.52% | 49.78% |
| 3,117 km$^2$ | 0.06% | 0.12% | 0.16% | 26.07% | 43.96% |
| 12,470 km$^2$ | 0.03% | 0.05% | 11.19% | 14.82% | 40.12% |
| 49,883 km$^2$ | 0.01% | 0.38% | 5.39% | 7.41% | 41.11% |
| Total CSES region | 1.21% | 1.69% | 2.44% | 4.43% | 26.11% |

For $M_t = 3$ and 3.5 (i.e., the first and second columns of Figure 4 across all spatial resolutions) and for the entire CSES region (i.e., the last row of Figure 4 across all $M_t$), the CIG curves of $ETAS_{\mu(x, y)}$ over $ETAS_\mu$ remain consistently above zero and exhibit an overall upward trend over time. This indicates that of $ETAS_{\mu(x, y)}$ outperforms the baseline $ETAS_\mu$ model in terms of forecasting skill under these settings. For other combinations of spatial resolution and $M_t$, however, the results are more ambiguous. The CIG curves often display substantial fluctuations (see discussion in Section 5.2), and in many cases, the performance of $ETAS_{\mu(x, y)}$ does not significantly exceed that of $ETAS_{\mu}$, reflecting the complex nature of seismicity in the study region. As with the parameter time series discussed in Section 4.1, many abrupt changes in CIG coincide with major earthquakes in the region, most notably the 2008 Wenchuan $M_S$ 8.0 earthquake (May 12), which is discussed in detail in Section 5.1.

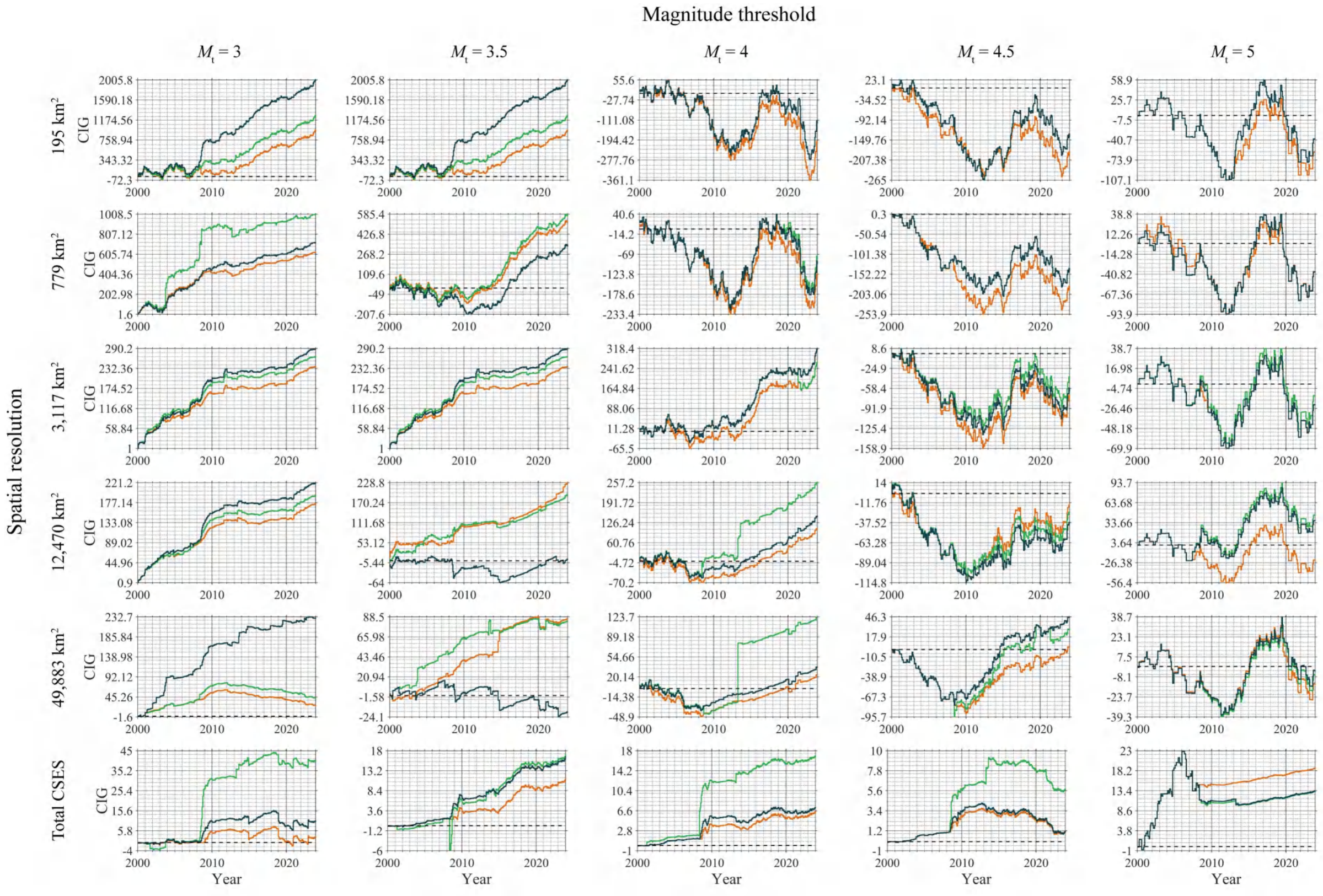

**Figure 4**. Time series of cumulative information gain (CIG) of $ETAS_{\mu(x, y)}$ over $ETAS_{\mu}$ in pseudo-prospective forecasting experiments at the CSES from 2000 to 2023. Results are shown for six spatial resolutions, five magnitude thresholds ($M_t$), and three strategies for assigning $P(n_i)$ when the observed value $n_i$ in the $i$-th triangle exceeds the forecasted maximum. The green, black, and orange curves correspond to strategies (1), (2), and (3) described in Figure 1, respectively. The pseudo-prospective forecasting experiments use calendar months as time

windows, resulting in a total of 288 time windows over the study period. The horizontal black dashed line indicates zero cumulative information gain.

**Alt-Text** Multi-panel time series showing the cumulative information gain (CIG) of the ETAS model with spatially variable background seismicity, $ETAS_{\mu(x,y)}$, relative to the constant-background model, $ETAS_{\mu}$, in pseudo-prospective monthly forecasts for the China Seismic Experimental Site from 2000 to 2023. Results are presented for six spatial grid resolutions and five magnitude thresholds ($M_t$). For cases in which the observed earthquake count in a spatial cell exceeds the simulated forecast maximum, three probability-assignment strategies are compared: green curves represent power law tail extrapolation, black curves represent a fixed small probability assignment, and orange curves represent exclusion of the probability term. Forecasts are evaluated over 288 monthly time windows. The horizontal black dashed line marks zero cumulative information gain, separating periods of positive and negative relative performance.

Figure 4 also shows that the choice of strategy for assigning $P(n_i)$ when the observed value $n_i$ exceeds the forecasted maximum can significantly influence the model evaluation results. For lower magnitude thresholds (e.g., $M_t = 3$ to 4), corresponding to the left-hand panels in Figure 4, the three strategies, power law extrapolation, fixed minimum probability, and exclusion, yield markedly different CIG curves. This highlights the sensitivity of model evaluation to how such rare cases are handled. In most instances, the power law extrapolation strategy results in the highest CIG values. However, there are exceptions, for example, under the combinations of spatial resolution 195, 3117, 12,470, and 49,883 km$^2$ with $M_t = 3$, the fixed minimum probability strategy consistently produces higher CIG curves than the power law extrapolation approach.

4.3 Bias from improper tail treatment

However, the apparent advantage of the CIG curves obtained using the fixed minimum probability strategy for assigning $P(n_i)$ is in fact illusory. When both models ($ETAS_{\mu(x,y)}$ and $ETAS_{\mu}$) assign $P(n_i)$ based on the same fixed minimum probability, i.e., when the observed earthquake count exceeds the maximum simulated count of both models, their log-likelihood scores become identical, resulting in an information gain (IG) of zero. In contrast, when the observed count exceeds the maximum simulated value of only one model but not the other, the former adopts the fixed minimum probability, while the latter still uses the adaptive Gaussian kernel density estimate to assign $P(n_i)$. This asymmetric treatment introduces a strong bias in model performance. Such cases frequently occur in the months following large earthquakes (e.g., events with $M > 7$); examples are observed for $M_t = 3.0$ during June and November 2008 in the full CSES region (see Data and Resources section).

During these post-mainshock forecast windows, $ETAS_{\mu(x,y)}$ systematically generates more tail events in the probability distribution $P(n)$ than $ETAS_\mu$. This difference arises because the spatially varying $ETAS_{\mu(x,y)}$ incorporates a heterogeneous background rate $\mu(x, y)$, which accounts for spatial clustering and variability in triggering intensity. Statistically, this produces a mixture of local Poisson-branching processes with different means, naturally leading to a fatter-tailed distribution. Physically, high-$\mu$ regions act as persistent triggering hotspots, sustaining multiple generations of aftershocks and resulting in abnormally large event counts within certain cells. Consequently, $ETAS_{\mu(x,y)}$ exhibits a broader and more asymmetric $P(n_i)$, consistent with the enhanced spatial heterogeneity of real seismicity.

In fact, the exclusion strategy for assigning $P(n_i)$ suffers from a similar problem, though its impact on CIG is less pronounced than that of the fixed minimum probability. Specifically, when both models assign probabilities under the exclusion rule (i.e., when the observed count exceeds the simulated maximum of both models), the IG of $ETAS_{\mu(x,y)}$ over $ETAS_\mu$ is zero. When the observed count exceeds the simulated range of only one model, the log-likelihood (*LL*) for that model cannot be computed, and thus the information gain (IG) for this forecast window becomes undefined. In practice, this window is omitted from the IG calculation, which explains why its impact on the overall CIG is less pronounced than that of the fixed minimum probability approach. In either case, whether using the fixed minimum probability or exclusion, the model performance is systematically biased following large earthquakes. Therefore, neither method should be recommended for evaluating IG in such contexts.

4.4 Spatial patterns of forecast performance

To further examine the spatial dependence of forecast performance, we mapped the mean information gain (MIG) of $\mathrm{ETAS}_{\mu(x,y)}$ over $\mathrm{ETAS}_{\mu}$ across all 288 monthly forecast windows from 2000 to 2023 (Figure 5). The analysis was performed at the finest spatial resolution considered in the present study, corresponding to triangular cells of approximately 195 km$^2$. For each cell, the information gain was first calculated for each forecast window as the log-likelihood difference between $\mathrm{ETAS}_{\mu(x,y)}$ and the benchmark $\mathrm{ETAS}_{\mu}$ model, and then averaged over the full study period. In calculating the information gain, out-of-range observations were handled using the power law extrapolation strategy, which has been shown in Sections 4.2 and 4.3 to provide the most theoretically justified and empirically robust treatment of the tail of the forecasted probability distribution $P(n)$. Therefore, Figure 5 provides a spatially resolved summary of the relative forecasting skill of $\mathrm{ETAS}_{\mu(x,y)}$ compared with $\mathrm{ETAS}_{\mu}$, after accounting for the fat-tailed nature of earthquake-count distributions, which is discussed in detail in Section 5.3.

The resulting map (and Figure 2) shows that the regions where $\mathrm{ETAS}_{\mu(x,y)}$ achieves positive mean information gain are not randomly distributed. Instead, they are spatially concentrated in areas that experienced earthquakes with $M \geq 5$ during the study period. These positive-gain regions indicate locations where the spatially heterogeneous background model provides a better description of the observed earthquake-count distribution than the homogeneous benchmark model. In contrast, regions without such moderate-to-large earthquakes generally exhibit weaker or more neutral differences between the two models. This spatial correspondence suggests that the advantage of $ETAS_{\mu(x, y)}$ is most evident in areas where seismicity is sufficiently active and clustered to provide informative constraints on the spatial variability of background seismicity and triggering behavior.

This pattern can be understood from both statistical and physical perspectives. Statistically, the spatially variable background rate $\mu(x, y)$ allows $ETAS_{\mu(x, y)}$ to redistribute long-term seismicity rates according to the observed spatial heterogeneity of the catalog, whereas $ETAS_{\mu}$ imposes a spatially uniform background rate over the entire region. As a result, in cells located near repeatedly active fault zones or near areas affected by $M \geq 5$ earthquakes and their aftershock sequences, $ETAS_{\mu(x, y)}$ can assign more realistic probabilities to elevated earthquake counts. Physically, these regions correspond to persistent or recurrent seismogenic structures where background activity, local triggering, and aftershock cascades are strongly localized in space. The heterogeneous background field therefore acts as a mechanism for incorporating such spatially persistent seismicity, improving the forecasted probability density functions $P(n)$ in active regions. By contrast, in relatively quiet areas, the number of observed events is small and the forecast evaluation is more strongly affected by stochastic fluctuations, making the performance difference between the two models less systematic.

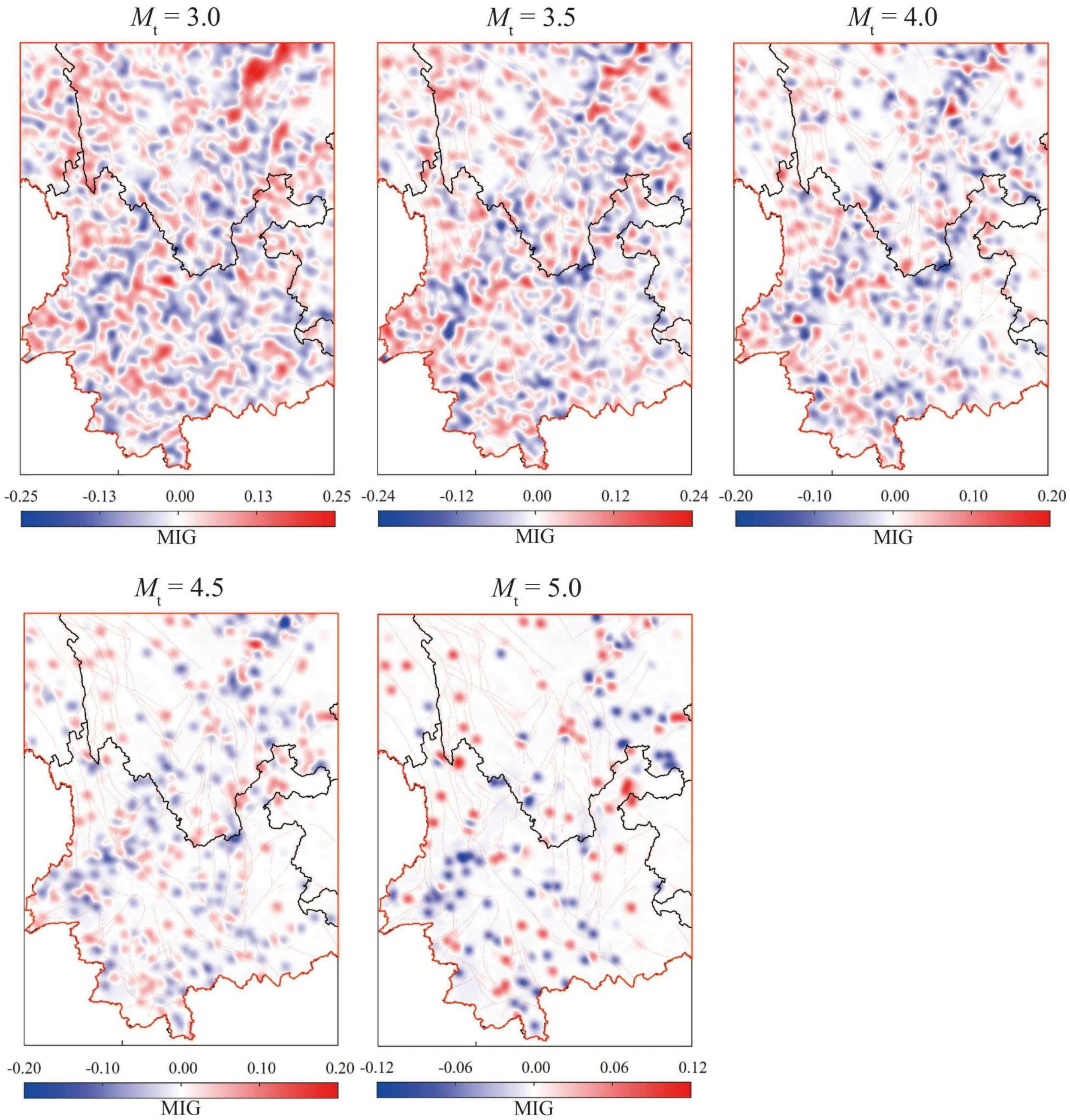

**Figure 5**. Mean information gain (MIG) of $ETAS_{\mu(x, y)}$ over $ETAS_{\mu}$ across a total of 288 time windows during the study period in pseudo-prospective forecasting experiments for the CSES from 2000 to 2023, at a spatial resolution of 195 km$^2$. The information gain used to compute the mean information gain is based on the power law extrapolation method. Red colors in the color bar indicate positive mean information gain, meaning that $ETAS_{\mu(x, y)}$ outperforms $ETAS_{\mu}$.

**Alt-Text** Map of the China Seismic Experimental Site divided into 195 square kilometer

triangular cells, showing the mean information gain of $\text{ETAS}_{\mu(x,y)}$ relative to $\text{ETAS}_{\mu}$ over 288 monthly forecast windows from 2000 to 2023. Positive values cluster mainly in areas where magnitude 5 or larger earthquakes occurred during the study period, indicating stronger performance of the heterogeneous-background model there; near-zero or negative values dominate quieter regions.

The 2008 Wenchuan sequence also raises the question of whether a single extreme cluster can disproportionately control the overall model ranking. We note that the present study does not rely solely on a single terminal 25-year score, but presents the forecast evaluation as CIG time series in Figure 4. This representation makes the contribution of large earthquake sequences explicit and temporally localized: the influence of the Wenchuan sequence can be identified from the abrupt change in the CIG curves around 2008, while the CIG increments before and after this period remain separately interpretable. Moreover, large earthquake sequences should not be regarded simply as outliers to be removed from forecast evaluation. They are precisely the periods during which short-term earthquake forecasting is most important, and the ability of a model to assign reasonable probabilities to aftershock-rich windows is an essential component of forecasting skill. Excluding the most extreme windows would therefore provide a conditional evaluation of model performance during relatively quiet periods, but would no longer assess the full pseudo-prospective forecasting problem considered here. The spatial pattern of mean information gain in Figure 5 further shows that the positive performance of $\text{ETAS}_{\mu(x,y)}$ is concentrated mainly in regions where $M \geq 5$ earthquakes occurred during the study period, indicating that the gain reflects physically meaningful clustered seismicity rather than an arbitrary artifact of a single extreme observation.

## 5 Discussion

In this section, we present the results of pseudo-prospective experiments conducted for May 2008 across the entire CSES region to quantitatively examine the impact of out-of-range earthquake counts on the calculation of information gain. Following this, we further explain the underlying causes of noisy cumulative information gain, highlighting both methodological and data-related factors. Finally, we examine the empirical scaling behavior of earthquake-count distributions and discuss its consistency with ETAS-based branching-process theory, including its implications for constraining the fertility exponent $\alpha$.

### 5.1 An example of out-of-range earthquake counts: The May 2008 Wenchuan earthquake

Within the spatio-temporal window of the present study, the largest event is the $M_S$ 8.0 Wenchuan earthquake of May 12, 2008. This mainshock and its aftershocks produced an out-of-range situation, as illustrated in Panel (b) of Figure 1, under all magnitude thresholds ($M_t$) considered in the present analysis (see Figure 6). Figure 6 also presents the probability distributions of earthquake counts forecasted by $ETAS_{\mu}$ and $ETAS_{\mu(x, y)}$ for the same month (May 2008) across the entire CSES region. The results clearly demonstrate that, once the observed counts fall outside the simulated ensemble, the extrapolation based on adaptive Gaussian kernel density estimation becomes severely biased and drastically underestimates the probability of the observed earthquake numbers. In contrast, power law distributions with fat-tailed behavior provide a more reliable extrapolation, yielding non-negligible probabilities for these extreme counts.

From Figure 6, the information gain (IG) of $ETAS_{\mu(x, y)}$ relative to $ETAS_{\mu}$ can be computed using Equation (3). The corresponding IG values are approximately 2, -2, -1, -3, and 1 for $M_t$ = 3, 3.5, 4, 4.5, and 5, respectively. These results suggest that, at the time of the Wenchuan mainshock itself, the two models did not exhibit a consistently large performance gap across thresholds. However, the bottom row of Figure 4 reveals a critical point: when extrapolation is performed using power law tails, the IG of $ETAS_{\mu(x,y)}$ relative to $ETAS_{\mu}$ continues to rise in the months following May 2008. This indicates that $ETAS_{\mu(x,y)}$ systematically outperforms $ETAS_{\mu}$ in forecasting the Wenchuan aftershocks, leading to the marked upward shift observed in the cumulative information gain (CIG) curve in Figure 4.

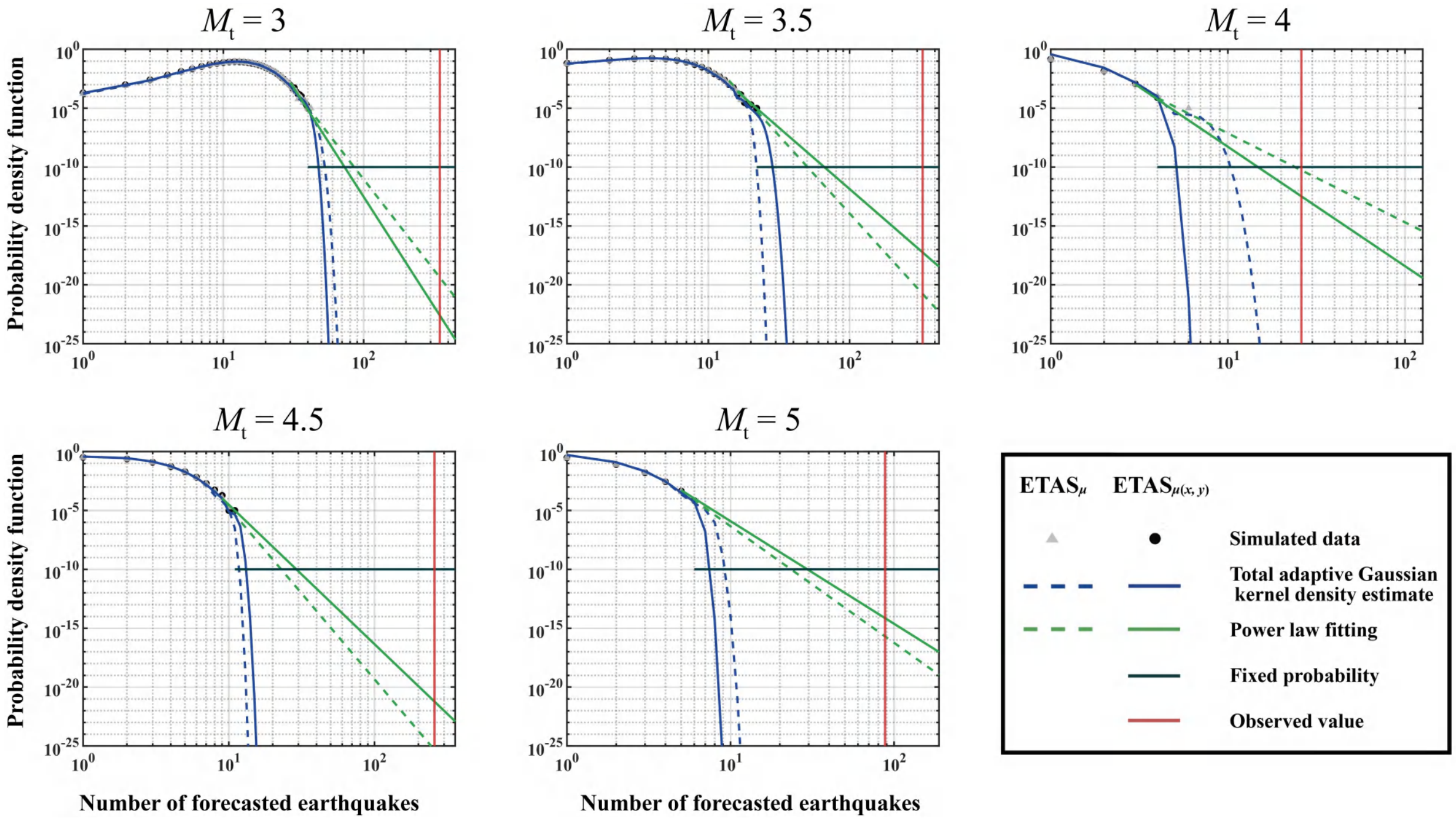

**Figure 6**. Probability distributions of earthquake counts forecasted by (dashed line) $ETAS_{\mu}$ and (solid line) $ETAS_{\mu(x,y)}$ for May 2008 across the entire CSES region, under different magnitude thresholds ($M_t$). The $M_S$ 8.0 Wenchuan earthquake on May 12 and its aftershocks (marked by the red vertical line) generate an out-of-range situation as shown in Panel (b) of Figure 1, which corresponds to a single time-window slice taken from the last row of Figure 4. All plotting elements are consistent with those in Figure 1, except that the $ETAS_{\mu}$ data points are shown as solid gray triangles for visual distinction. The Data and Resources section provides links to access the forecasted probability density functions of both models for all forecast windows and magnitude thresholds across the full CSES region.

**Alt-Text** Probability density functions of forecasted earthquake counts for May 2008 across the full China Seismic Experimental Site (CSES), comparing the ETAS model with constant background rate, $ETAS_{\mu}$ (dashed curve), and the spatially variable background model, $ETAS_{\mu(x,y)}$ (solid curve), under multiple magnitude thresholds ($M_t$). Simulated forecast samples are displayed as symbols, with ETAS$_{\mu(x,y)}$ shown as black circles and ETAS$_\mu$ shown as solid gray triangles. The observed earthquake count, dominated by the 12 May 2008 $M_S$ 8.0 Wenchuan earthquake and its aftershocks, is indicated by a red vertical line and lies beyond the maximum simulated range, illustrating the out-of-range case described in panel (b) of Figure 1. Kernel-based density estimates and plotting conventions follow those in Figure 1. Links to the full set of forecasted probability density functions for all time windows and magnitude thresholds are provided in the Data and Resources section.

To examine whether the choice of the upper 30% tail fraction introduces a substantial bias in the power law extrapolation, we performed a sensitivity test by repeating the extrapolation using upper-tail fractions from 10% to 50% of the simulated data (Table 2). This test is intended to assess the stability of the extrapolated likelihood for the May 2008 out-of-range case, rather than to infer the theoretical scaling exponent $a$ discussed in Section 5.3. The results show that the 10% and 20% choices do not always provide stable estimates across all magnitude thresholds and both ETAS variants, as indicated by missing entries in several cases. This suggests that using too small a fraction of the simulated tail may leave insufficient support for a robust extrapolation, especially when the forecasted distribution becomes sparse at higher magnitude thresholds. In contrast, tail fractions of 30% to 50% provide finite estimates for all magnitude threshold values $M_t$ and for both ETAS$_\mu$ and ETAS$_{\mu(x,y)}$. Although the resulting $\log_{10}P(n_i)$ values vary with the chosen tail fraction, as expected for far-tail extrapolation, they consistently indicate extremely small but non-zero probabilities for the May 2008 out-of-range observations. The qualitative conclusion therefore remains unchanged: the Wenchuan mainshock and its aftershocks generated earthquake counts far beyond the simulated range, and power law extrapolation provides a more appropriate treatment of such extreme observations than either assigning a fixed minimum probability or excluding the corresponding likelihood term. We therefore adopted the upper 30% tail fraction as the default choice because it is the smallest fraction that yields complete estimates across all tested configurations, while still focusing on the upper part of the simulated distribution. This choice represents a practical bias-variance compromise and does not drive the main conclusions of the present study.

**Table 2.** Sensitivity of power law extrapolation to the upper-tail fraction used for power law fitting. Values are $\log_{10}P(n_i)$, where $P(n_i)$ is the forecasted probability assigned to the observed earthquake count $n_i$ by $ETAS_\mu$ and $ETAS_{\mu(x,\,y)}$ for May 2008 across the entire China Seismic Experimental Site (CSES) region, under different magnitude thresholds $M_t$. The em dash indicates cases in which no stable extrapolated probability was obtained.

| Tail fraction | $M_t = 3$ | | $M_t = 3.5$ | | $M_t = 4$ | | $M_t = 4.5$ | | $M_t = 5$ | |
|---|---|---|---|---|---|---|---|---|---|---|
| | $ETAS_\mu$ | $ETAS_{\mu(x,\,y)}$ | $ETAS_\mu$ | $ETAS_{\mu(x,\,y)}$ | $ETAS_\mu$ | $ETAS_{\mu(x,\,y)}$ | $ETAS_\mu$ | $ETAS_{\mu(x,\,y)}$ | $ETAS_\mu$ | $ETAS_{\mu(x,\,y)}$ |
| 10% | -15.49 | -23.43 | — | -13.79 | — | — | — | — | — | — |
| 20% | -18.04 | -18.43 | -21.87 | -10.72 | -29.21 | — | — | — | — | — |
| 30% | -19.37 | -17.28 | -19.73 | -21.34 | -22.88 | -23.46 | -14.69 | -17.49 | -9.28 | -13.03 |
| 40% | -16.07 | -15.30 | -16.45 | -19.67 | -22.38 | -20.38 | -15.63 | -15.56 | -11.07 | -12.81 |
| 50% | -12.30 | -12.10 | -13.74 | -16.23 | -15.97 | -18.19 | -16.31 | -13.22 | -11.07 | -10.32 |

5.2 Quantitative rationale for the variability of CIG

Figure 4 shows that the CIG curves of $ETAS_{\mu(x,\,y)}$ relative to $ETAS_\mu$ lose their long-term upward trend and become increasingly erratic when the spatial resolution is coarsened and $M_t$ is raised. Below we offer a quantitative rationale for this behaviour.

Let $n$ be the number of earthquakes observed in a single grid cell at the chosen resolution $r$ (expressed as the mesh area) during the $i$th forecast window for which there are $N_r$ grid cells. The number of grid cells scales as $N_r \sim 1/r$. Let us assume that $n$ follows a discrete probability law $P(n)$ with mean $n_0$ (> 0). Suppose models A and B also generate counts from $P(n)$ but with means $n_A$ and $n_B$, respectively, subject only to the accuracy condition $|n_A - n_0| < |n_B - n_0|$ indicating that model A is better than model B. Define the per-cell log-likelihood difference during $i$th forecast window

$$\Delta LL_i = \log P_A^i(n) - \log P_B^i(n) \tag{6}$$

Since $|n_A - n_0| < |n_B - n_0|$, the average log-likelihood difference Mean($\Delta LL$) is positive, indicating that model A is the better fit. The accuracy of this statistic is assessed with the coefficient of variation (CV) of the log-likelihood difference:

$$\mathrm{CV} = \frac{\mathrm{Std}(\Delta LL)}{\mathrm{Mean}(\Delta LL)} \tag{7}$$

where the mean and standard deviation are calculated over the set of $N_r$ cells. The CV is a function of the resolution $r$ via the dependence of the number of grid cells $N_r \sim 1/r$. The coarser the resolution $r$ is, the larger is $n_0$ and the narrower is $P(n)$, but the smaller is $N_r$. Thus coarsening the resolution $r$ improves the statistics per cell but deteriorates the statistical accuracy obtained by averaging over all $N_r$ cells. This competition between in-cell and ensemble-cell statistics explains the complicated dependences of the CIG curves on the resolution and magnitude threshold shown in Figure 4.

Table 3 reports the CV computed solely from the orange CIG curves in Figure 4, i.e., using the exclusion strategy (see more details in Section 2.3). Under this strategy only the probabilities obtained from the adaptive-Gaussian kernels are retained; any cells whose observed counts fall outside the simulated range are discarded. This ensures that all CV values are derived from a single, internally consistent method, without the complications introduced by fixing minimum probability or power law extrapolation. The results show a clear pattern: for a given spatial resolution, a larger $M_t$ lowers the average count $n_0$ (e.g., Figures 6 and 7); the mean of $\Delta LL$ therefore declines while its sampling variance grows, and the CV increases accordingly.

The CV values in Table 3 should be interpreted as an empirical signal-to-noise diagnostic for the stability of the CIG curves, rather than as a formal hypothesis test. In general, configurations with small CV values, approximately CV < ~10, show relatively stable CIG behavior and are therefore more robustly interpretable. Configurations with intermediate CV values, roughly between 10 and 20, should be interpreted together with the corresponding CIG time series in Figure 4, because some of them still exhibit coherent long-term trends despite moderate variability. For example, several low-threshold configurations with $M_t = 3$ or $M_t = 3.5$ display persistent positive CIG trends even when their CV values are slightly larger than 10. In contrast, configurations with CV values well above 20 are increasingly dominated by stochastic variability, and extreme cases, such as CV = 93.82 for $M_t = 4.5$ at 12,470 km$^2$, should not be interpreted as providing reliable evidence for systematic model superiority or inferiority. Overall, Table 3 and Figure 4 jointly indicate that statistically interpretable CIG results are mainly obtained for data-rich settings, especially lower magnitude thresholds and configurations with persistent CIG trends, whereas high-$M_t$ and sparse-count settings are largely noise dominated.

**Table 3.** Coefficient of variation (CV) calculated based on the cumulative information gain (CIG) results for ETAS$_{\mu(x,\ y)}$ relative to ETAS$_{\mu}$ in the China Seismic Experimental Site (CSES) region, evaluated for each combination of spatial resolution and magnitude threshold $M_t$. The CV values are calculated from the Exclusion strategy only, that is, from the orange CIG curves shown in Figure 4.

| Spatial resolution | $M_t = 3$ | $M_t = 3.5$ | $M_t = 4$ | $M_t = 4.5$ | $M_t = 5$ |
|---|---|---|---|---|---|
| 195 km$^2$ | 7.28 | 15.73 | 22.92 | 16.02 | 34.62 |
| 779 km$^2$ | 3.87 | 7.19 | 28.18 | 13.38 | 48.38 |
| 3,117 km$^2$ | 3.08 | 4.64 | 9.05 | 27.25 | 68.67 |
| 12,470 km$^2$ | 2.62 | 3.42 | 13.03 | 93.82 | 54.27 |
| 49,883 km$^2$ | 10.90 | 4.72 | 25.78 | 21.50 | 64.33 |

### 5.3 Consistency of empirical scaling and ETAS-based theoretical predictions

Figure 7 demonstrates that the empirical probability density functions (PDFs) of earthquake counts per grid cell are well approximated by the power law $P(n) \sim n^{-(1+a)}$ across all tested magnitude thresholds and spatial resolutions. In Figure 7, the scaling behavior is further examined separately for the full study period, the pre-Wenchuan period, and the post-Wenchuan period, in order to evaluate whether the 2008 Wenchuan $M_S$ 8.0 earthquake produces a systematic change in the fitted exponent. For each combination of spatial resolution and magnitude threshold, the mean and standard deviation of the fitted exponent *a* were estimated from 200 bootstrap resamples. Figure 8 then plots the histogram of the 25 mean *a* values shown in Figure 7 together with their Gaussian fit for the full study period, the period before Wenchuan, and the period after Wenchuan, respectively. The corresponding ensemble-averaged exponents are $a_t = 1.02 \pm 0.23$, $a_b = 1.47 \pm 0.45$, and $a_a = 0.92 \pm 0.22$, respectively. We therefore regard $a_t = 1.02 \pm 0.23$ as an effective ensemble-averaged exponent across spatial resolutions and magnitude thresholds, rather than as an exact universal exponent for every individual configuration.

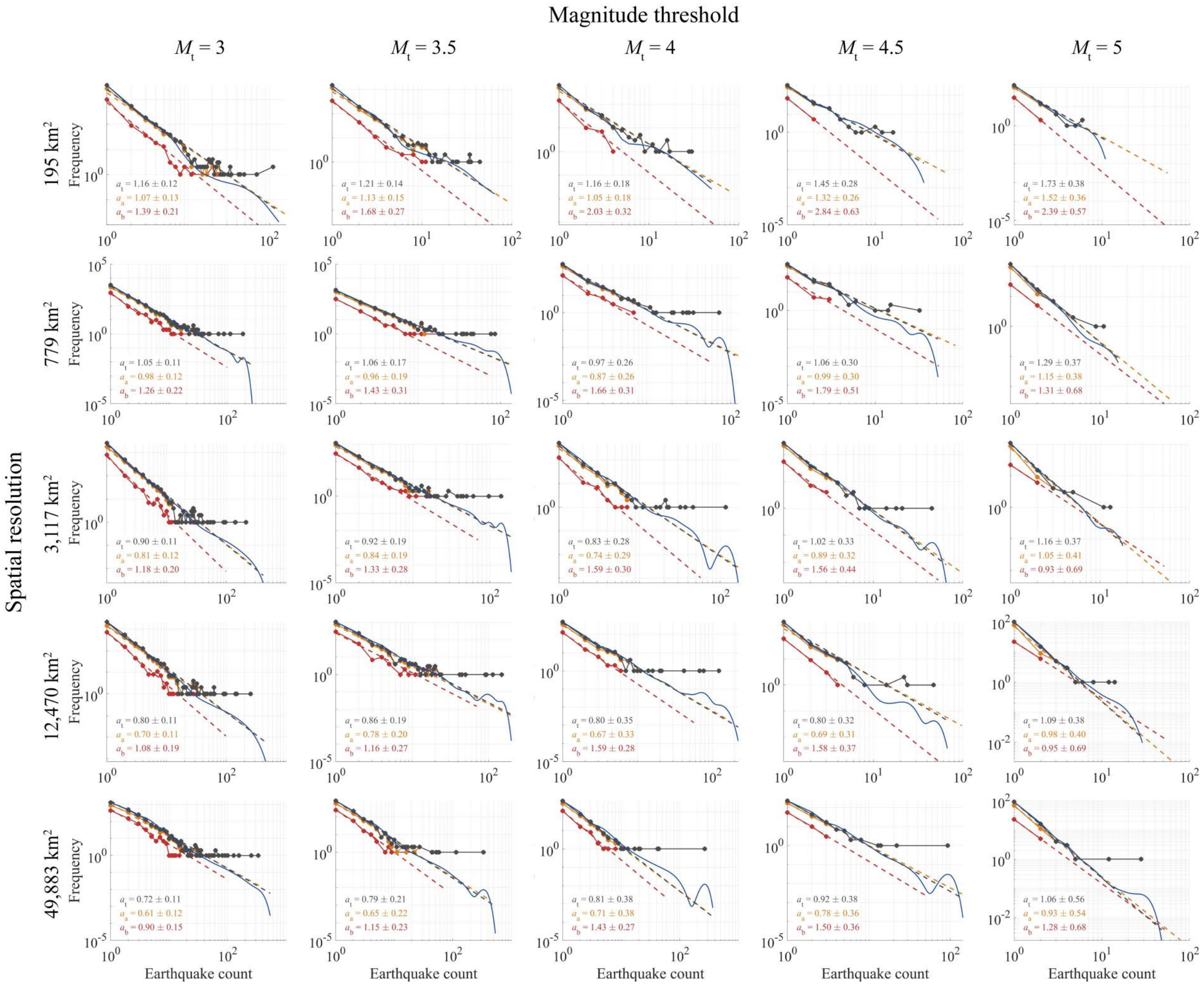

**Figure 7**. Frequency distribution function of earthquake count per grid cell over the ensemble of grid cells, aggregated over all 288 forecast windows for the China Seismic Experimental Site. Each point represents the number of grid cells (on the $y$-axis) that experienced a specific number of earthquakes (on the $x$-axis), for a specific spatial resolution and magnitude threshold. In other words, the curves show the empirical non-normalized probability density function of earthquake counts per cell. The black, orange, and red fitted lines denote the power law fits for the full study period, the pre-Wenchuan period, and the post-Wenchuan period, respectively. The corresponding exponents are denoted by $a_t$, $a_b$, and $a_a$. Consistent with Figure 1, the blue solid line represents the adaptive Gaussian kernel density estimate, while the dashed lines show the fitted power law function ($P(n) \sim n^{-(1+a)}$). For each combination of spatial resolution and magnitude threshold, the mean and standard deviation of the fitted exponent $a$ were estimated from 200 bootstrap resamples.

**Alt-Text** Frequency distributions of earthquake counts per grid cell, aggregated over 288 monthly forecast windows for the China Seismic Experimental Site. Each point represents the number of spatial cells (vertical axis) that experienced a given earthquake count (horizontal axis), for a specific spatial resolution and magnitude threshold. The curves therefore depict the empirical, non-normalized probability density of earthquake counts across cells. The blue solid curve shows the adaptive Gaussian kernel density estimate of the distribution, and the dashed curves show a fitted power law function proportional to $n^{-(1+a)}$. For each spatial resolution and magnitude threshold, the exponent $a$ and its standard deviation were estimated using 200 bootstrap resamples.

Using the technique of generating probability functions applied to the ETAS model, Saichev et al. (2005) derived the theoretical distribution of the total number of aftershocks, which follows the power law $P(n) \sim n^{-(1+a)}$ in the non-critical regime (i.e., when its branching ratio $h$ is not close to 1 and for earthquake counts $n > 1/(1 - h)^{g/(g-1)}$ where $g = b/\alpha$, $b$ is the Gutenberg-Richter (GR) slope and $\alpha$ is the fertility (or productivity) exponent describing how the mean number of aftershocks triggered by a mainshock scales with its magnitude. The power law exponent $a$ is related to the model parameters through $a = g = b/\alpha$. For smaller event counts $n < 1/(1 - h)^{g/(g-1)}$, the exponent $a$ is renormalized to $a = 1/g = \alpha/b$. The empirical exponents obtained here are mostly close to or larger than unity within uncertainty, suggesting broad consistency with the branching-process scaling regime, although they should not be interpreted as a strict diagnostic of criticality in finite and non-stationary observational windows. This finding corroborates previous studies suggesting that the Earth's crust, on average, remains far from criticality (Nandan et al., 2021b; Li et al., 2025a). Saichev & Sornette (2006) further demonstrated that this same scaling law emerges as the asymptotic behavior of space-time windowed event-number distributions, implying that such power laws are a fundamental statistical property of branching-type seismicity models.

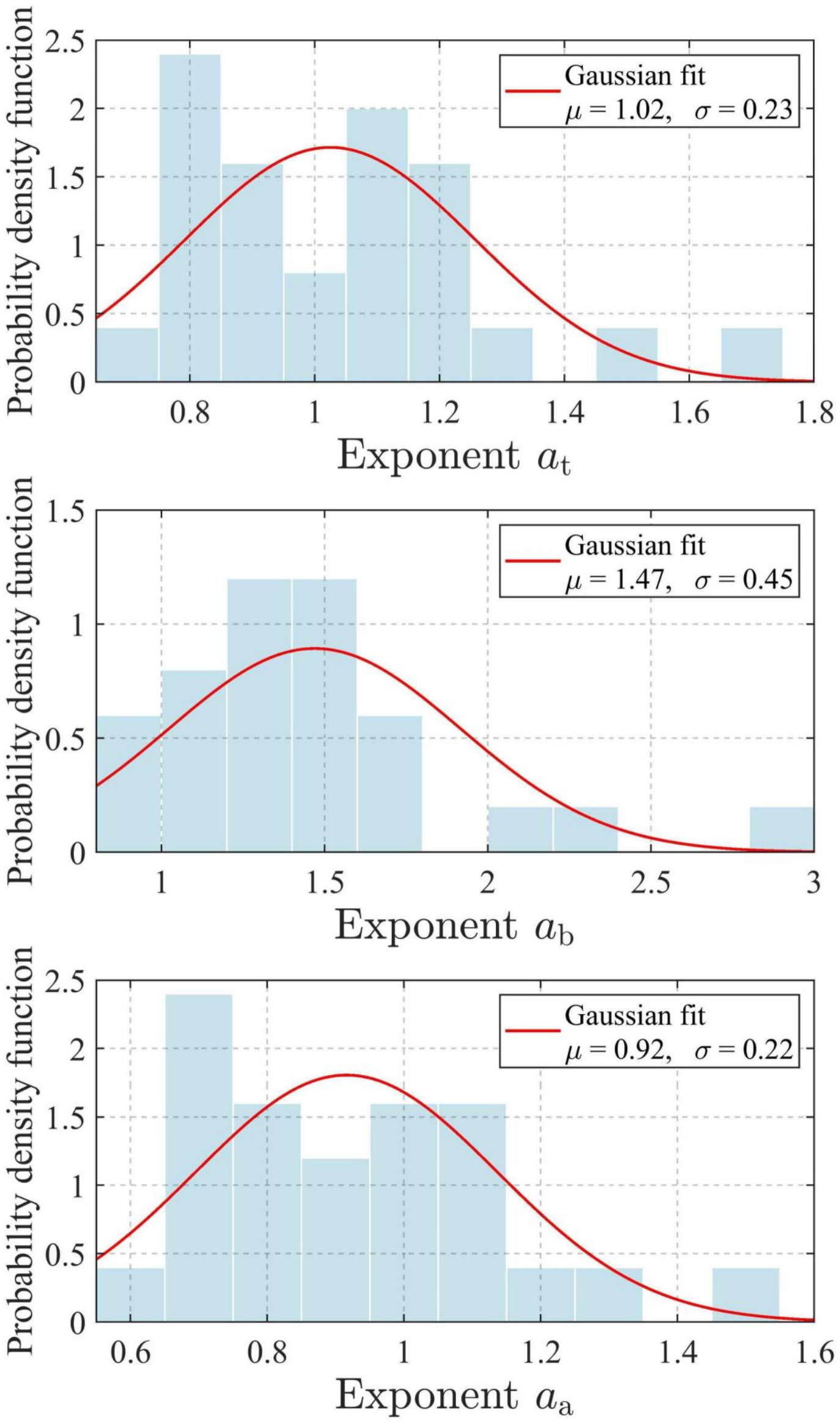

**Figure 8**. Histograms of the 25 mean power law exponents *a* shown in Figure 7 for the total (top panel), pre- (middle panel) and post- (bottom panel) Wenchuan 2008 earthquake and their Gaussian fit.

**Alt-Text** Three vertically stacked histograms showing the distributions of 25 mean power law exponents derived from the earthquake count distributions in Figure 7. Each panel includes a Gaussian fit shown by the red line.

We emphasize, however, that the theoretical result of Saichev et al. (2005) strictly applies to the total progeny of an idealized stationary branching process in unbounded space and time. In contrast, the empirical distributions analyzed here are constructed from earthquake counts within finite one-month forecast windows, finite spatial cells, and fixed magnitude thresholds, in a region that exhibits temporal non-stationarity, particularly around the 2008 Wenchuan earthquake. Therefore, the comparison between theory and data should be interpreted in an approximate asymptotic sense, rather than as a strict finite-sample validation of the idealized theory. The relevance of the theory in the present setting is supported mainly by the extension of Saichev & Sornette (2006, 2007), who showed that similar power law scaling can emerge for event-number distributions measured in finite space-time windows. Under realistic observational constraints, finite temporal windows, spatial truncation, boundary effects, and magnitude thresholds are expected to distort the distribution, especially at small earthquake counts. Accordingly, the theoretical scaling is expected to be most meaningful in the intermediate-to-large count regime, which is the part of the distribution used for tail fitting in the present study.

Figure 7 also provides a direct check on the possible influence of the 2008 Wenchuan earthquake. Although local differences between the pre- and post-Wenchuan fitted exponents are visible in some panels, the two sets of estimates remain broadly overlapping within their uncertainties determined by bootstrap. The pre-Wenchuan distribution yields a larger ensemble-averaged exponent, $a_b = 1.47 \pm 0.45$, whereas the post-Wenchuan distribution gives $a_a = 0.92 \pm 0.22$. This contrast points to a possible modification of the effective scaling behavior by the Wenchuan sequence and its aftermath; however, the overlap between the uncertainty ranges suggests that the two exponents may be statistically indistinguishable at this level of precision. Thus, the hypothesis that the pre- and post-Wenchuan distributions share a common scaling exponent cannot be rejected from these estimates alone. Instead, the Wenchuan earthquake represents a strong transient episode superimposed on a broader branching-process-like scaling behavior. We also note that temporal variations in magnitude completeness, especially immediately after strong earthquakes, may affect the observed event counts. However, the persistence of similar scaling behavior across multiple magnitude thresholds, together with the catalog-completeness analysis discussed in Section 3, suggests that short-term catalog incompleteness does not dominate the inferred scaling.

A systematic dependence of the fitted exponent $a$ on the magnitude threshold $M_t$ is nevertheless visible in Figure 7: lower magnitude thresholds tend to produce smaller fitted exponents, whereas higher magnitude thresholds tend to yield larger values. We interpret this trend primarily as an effective finite-sample and thresholding effect rather than as evidence against the branching-process scaling relation. Increasing $M_t$ reduces the number of events available in each space-time bin and shifts the fitted distribution toward sparser and more discrete count statistics. At the same time, higher thresholds preferentially retain larger events and their more localized aftershock productivity, so that the fitted tail can reflect a threshold-dependent effective exponent. Conversely, at lower $M_t$, the larger number of small events provides better sampling of the full cascade and tends to broaden the count distribution, leading to smaller fitted $a$. Thus, the full-period value $a_t = 1.02 \pm 0.23$ should be viewed as a compact summary of the ensemble behavior across all tested configurations, while the threshold-dependent deviations reflect finite-window sampling, magnitude-threshold effects, and residual non-stationarity in the observed catalog.

Moreover, the observed power law exponent $a$ provides an indirect way to infer the ETAS fertility (or productivity) exponent $\alpha$. Given the theoretical relation $a = g := b/\alpha$, the empirical result $a_t = 1.02 \pm 0.23$ and the Gutenberg-Richter slope value $b = 0.85 \pm 0.002$ (see Li et al., 2025a) yield an estimated fertility exponent of $\alpha_t = b/a_t = 0.83 \pm 0.19$ for the full study period. Applying the same error-propagation calculation separately to the pre- and post-Wenchuan exponents gives $\alpha_b = 0.58 \pm 0.18$ and $\alpha_a = 0.92 \pm 0.22$, respectively. Here, the uncertainty of $\alpha$ was calculated by standard error propagation, assuming that the uncertainties in $a$ and $b$ are independent. Although the uncertainty in $b$ is relatively small compared with the variability in $a$, including it provides a more realistic estimate, especially given the mixed magnitude types in the CSES catalog. This inferred value of $\alpha_t$ falls well within the physically plausible range reported by previous ETAS-based studies and reinforces the interpretation that seismic triggering efficiency increases moderately with magnitude but remains close to the marginal regime ($\alpha \approx b$). The parameter estimates obtained directly from the ETAS-based models give (Figure 3): $\alpha \approx 0.7$–$1.0$ for $\text{ETAS}_{\mu(x,y)}$ and $\alpha \approx 0.6$–$0.8$ for $\text{ETAS}_{\mu}$, in good agreement.

Although the indirect estimate inferred from the count-distribution scaling is in good agreement with the directly calibrated ETAS parameters, the two estimates are not expected to be identical. Several factors may account for the remaining differences between them. First, the theoretical derivation assumes an idealized, stationary branching process in the non-critical regime, while the ETAS fitting is affected by non-stationary background rates, spatial heterogeneity, and catalog incompleteness, all of which can modify the apparent fertility. In particular, undetected small earthquakes can break the true parent-offspring structure of the cascade. Their descendants may then be attributed to larger observed events, making large earthquakes appear more productive than they actually are and thereby increasing the estimated $\alpha$. This mechanism is closely related to the apparent-clustering bias caused by incomplete earthquake catalogs (Sornette & Werner, 2005b) and to the sensitivity of ETAS parameter estimation to short-term aftershock incompleteness (Omi et al., 2014; Zhuang et al., 2017; Mizrahi et al., 2021; Hainzl, 2022; Hainzl et al., 2024). Second, the estimation of $\alpha$ in ETAS models is strongly coupled with that of other parameters, particularly the background rate $\mu$ and the spatio-temporal decay parameters ($1+\rho$ and $1+\omega$), leading to potential trade-offs that shift the fitted values upward. For example, an underestimated background rate or an overly concentrated spatial triggering kernel can be compensated by a larger productivity exponent, because the likelihood can improve by assigning more offspring to larger events. Third, the boundary and finite-size effects inherent in the CSES region (Li et al., 2025a), especially the presence of clustered aftershock sequences such as for the 2008 Wenchuan earthquake, may enhance the observed triggering efficiency relative to the long-term theoretical expectation.

Despite these complications, the full-period estimate $a_t = 1.02 \pm 0.23$ remains broadly consistent with both the directly calibrated ETAS values and the Gutenberg-Richter slope $b = 0.85 \pm 0.002$. This indicates that the system lies close to the transition between small-event-dominated and large-event-dominated triggering. Therefore, rather than claiming a strong ultraviolet-divergence regime (Sornette & Werner, 2005a), our results support a more cautious interpretation: small earthquakes remain collectively important because of their large numbers, but the inferred productivity exponent is close enough to $b$ that finite-window effects, catalog incompleteness, and large aftershock sequences can substantially affect the apparent balance between small- and large-earthquake triggering. Because direct empirical estimation of the fertility exponent $\alpha$ is notoriously difficult, often hindered by limited temporal completeness and strong parameter correlations, our indirect inference offers a robust and independent constraint. It provides empirical support for theoretical expectations derived from branching process theory and anchors a key ETAS parameter on a statistically grounded foundation directly linked to the observed scaling behavior of earthquake count distributions.

## 6 Conclusions

Advancing short-term earthquake forecasting has always been a core objective of the emerging China Seismic Experimental Site (CSES), led by China Earthquake Administration (CEA). A critical step in that agenda is the rigorous testing of scientific hypotheses on earthquake predictability. To that end, we have developed AoyuX, a bespoke short-term predictability platform tailored to the CSES region, which is currently in its initial implementation stage. AoyuX embeds the ETAS model, one of the few frameworks that simultaneously delivers time, space, magnitude, and probability information and that synthesises the current state of seismological knowledge. By embedding two variants of the ETAS model, $\text{ETAS}_{\mu}$ with a homogeneous background rate and $\text{ETAS}_{\mu(x,y)}$ with a spatially variable rate, we carried out about 25 years of monthly pseudo-prospective experiments (2000-2023).

The calibration phase reveals systematic contrasts between the two variants. $\text{ETAS}_{\mu(x,y)}$ consistently assigns more events to the background than $\text{ETAS}_{\mu}$, reflecting the added flexibility of a non-parametric background field. Several parameter time series exhibit abrupt shifts that coincide with major regional earthquakes, indicating that large events can instantaneously reshape the triggering environment as well as be associated with transient data issues such as earthquake incompleteness. Forecast assessments, expressed through cumulative information gain (CIG), show a clear hierarchy under data-rich conditions: for the full CSES domain and for $M_t = 3.0$ to 3.5, $\text{ETAS}_{\mu(x,y)}$ maintains a persistent positive CIG over $\text{ETAS}_{\mu}$. The spatial distribution of mean information gain further shows that the advantage of $\text{ETAS}_{\mu(x,y)}$ is concentrated mainly in regions where $M \geq 5$) earthquakes occurred during the study period, rather than being uniformly distributed over the entire CSES region. However, in sparse-count configurations, caused by fine spatial discretization, high magnitude thresholds, or limited seismicity in many cells, sampling variance grows and the CIG curves become highly volatile. The coefficient-of-variation analysis provides an empirical signal-to-noise diagnostic: low-CV configurations yield more stable and interpretable CIG curves, whereas very large CV values indicate that model comparison is dominated by stochastic variability rather than by coherent forecasting skill.

Model ranking also depends sensitively on how "out-of-range" cell counts, observed counts exceeding all simulated values, are handled. Among the strategies tested, power law extrapolation provides the most theoretically justified and empirically robust results, in line with both analytical predictions of branching-process theory and the clear scaling patterns shown in Figure 7 and earlier studies (Saichev et al., 2005; Saichev & Sornette, 2006; 2007). While in a few coarse-grained low-magnitude configurations, the fixed minimum probability rule produced slightly higher CIG values, these are best interpreted as artifacts of noise rather than genuine model superiority. Both the fixed minimum probability and exclusion strategies introduce systematic biases in model evaluation following large earthquakes. Consequently, these approaches are not suitable for reliable assessment of information gain during periods of intense seismic activity. Taken together, the evidence strongly supports power law extrapolation as the most appropriate framework for handling extreme counts for the ETAS-based branching models studied here, especially in cells affected by large, cluster-forming earthquakes, which generate the fat-tailed distributions expected from aftershock cascades. This highlights the necessity of treating such extreme outliers not as anomalies to be truncated but as integral signatures of the underlying triggering dynamics.

Although the present study focuses on ETAS-based forecasts, the proposed tail-aware scoring framework is not intrinsically tied to the ETAS model. Its essential requirement is that a forecasting model can provide, either analytically or through simulations, the full probability distribution $P(n)$ of earthquake counts in each space–time–magnitude bin. This condition can also be satisfied by machine learning-based forecasts, physics-based models, or hybrid approaches, provided that they generate probabilistic or ensemble forecasts rather than only deterministic point estimates. In such applications, however, the appropriate tail treatment should be diagnosed from the model-specific and data-specific count distributions. For ETAS-type branching models, power law extrapolation is theoretically motivated and empirically supported by the fat-tailed distributions observed here. Other model classes may produce different tail behavior, for example exponential, lognormal, or bounded distributions, in which case the extrapolation rule should be adapted accordingly. Therefore, the broader contribution of the present framework is not the exclusive use of a power law tail, but the explicit recognition that out-of-range observations must be handled by a transparent, empirically testable, and physically interpretable tail model. This makes the framework potentially useful for fair model comparison across ETAS, machine learning, physics-based, and future hybrid earthquake forecasting models.

As a complementary result, our analysis provides independent estimates of the fertility exponent inferred from the empirical scaling of earthquake-count distributions and the ETAS-based branching-process theory. For the full study period, we obtain $a_t = 1.02 \pm 0.23$, which, together with $b = 0.85 \pm 0.002$, yields $\alpha_t = 0.83 \pm 0.19$. The pre- and post-Wenchuan periods yield $\alpha_b = 0.58 \pm 0.18$ and $\alpha_a = 0.92 \pm 0.22$, respectively. These values should be interpreted as effective estimates under finite space-time windows and non-stationary catalog conditions, rather than as exact universal constants. Nevertheless, they provide useful independent constraints on this key but difficult-to-estimate ETAS parameter and indicate that the productivity exponent is close to the Gutenberg-Richter *b*-value in the CSES region.

In sum, $\mathrm{ETAS}_{\mu(x,\,y)}$ provides demonstrable skill over its homogeneous counterpart in regimes where observational counts are plentiful and seismicity is sufficiently active and clustered. Where data are sparse, however, sampling noise, parameter instability, and treatment of extreme clusters become dominant concerns. Future work should therefore focus on adaptive strategies for rare-event handling, model-specific tail diagnostics, and on hybrid schemes that combine ETAS-style triggering with physics-based constraints, further enhancing the operational value of AoyuX for short-term earthquake forecasting in the CSES and analogous natural laboratories worldwide.

## Acknowledgments

The authors would like to thank Yang Zang from the China Earthquake Networks Center for his assistance in accessing the Chinese earthquake catalog data, as well as Prof. Jiancang Zhuang, Prof. Zhongliang Wu and Prof. Changsheng Jiang for their valuable suggestions, Xinyi Wang for assistance in computing catalog completeness using the $\mathrm{GR_{AE}}$ method, and the two anonymous reviewers for their constructive comments and suggestions. This work is partially supported by the Guangdong Basic and Applied Basic Research Foundation (Grant No. 2024A1515011568) and the Center for Computational Science and Engineering at Southern University of Science and Technology.

## Data and Resources

The earthquake catalogs used in this study were obtained from the National Earthquake Science Data Center of China (https://data.earthquake.cn/datashare/report.shtml?PAGEID=earthquake_zhengshi), which is publicly accessible; however, user registration is required to access the catalog information. The catalog of earthquakes with magnitudes ≥ 3.0 in the CSES region during 1970–2023 is provided as Data Set S1 in Li et al. (2025a). The simulated earthquake catalogs generated by the two ETAS models can also be accessed through Li et al. (2025d). In addition, the forward monthly, quarterly, semi-annual, and annual AoyuX earthquake forecast reports for the CSES since June 2024 are available at Li (2025). Links to the forecasted probability density functions of both models for all forecast windows and magnitude thresholds across the CSES region are available at Li (2025a; 2025b; 2025c; 2025d; 2025e).

## Figure Legends

**Figure 1**. Illustration of the probability distribution for the number of forecasted earthquakes, shown for two scenarios: (a) where the observed value (red solid line) falls within the range of simulated values, and (b) where the observed value exceeds the maximum of the simulated distribution. The forecasted probability corresponding to the observed value $n_i$ in the $i$-th triangle is denoted as $P(n_i)$. In the present study, simulated data (black solid circles) are generated from $N_{\mathrm{sim}}$ sets of synthetic catalogs ($N_{\mathrm{sim}}$ = 100,000). The per-sample adaptive Gaussian kernels (gray shaded areas), each with its own bandwidth, are summed to construct the total adaptive Gaussian

kernel density estimate (blue solid line), which is used to present the overall probability distribution. When $n_i$ (red solid line) exceeds the forecasted maximum (orange dashed line) in panel (b), we propose three alternative strategies for assigning $P(n_i)$: (1) fitting a power law function (green solid line) to the upper 30% of the simulated tail data (to the right of the green dashed line); (2) assigning a fixed probability value (black horizontal line), equal to $1/(N_{\mathrm{sim}})^2$; or (3) excluding $P(n_i)$ altogether in such cases.

**Figure 2**. Spatio-temporal distribution of seismicity and completeness magnitude ($M_c$) in the China Seismic Experimental Site (CSES), located in the Sichuan-Yunnan region. (a) Spatial distribution of earthquakes from 1970 to 2023, completeness magnitude ($M_c$) since 2009, and active faults within the CSES. The background map of $M_c$ is adapted from Li et al. (2023). Fault locations are taken from Deng et al. (2003). (b) Location of the CSES shown on a broader regional map. (c) Temporal evolution of $M_c$ (right $y$-axis) and the magnitude-time (MT) plot of earthquakes with $M \geq 3$ (left $y$-axis) since 1985. The estimation of $M_c$ was performed using a half-year step and one-year window. Red dashed lines with decreasing color indicate completeness probabilities of 84.2% ($M_c + \sigma_c$), 97.8% ($M_c + 2\sigma_c$), and 99.9% ($M_c + 3\sigma_c$), respectively. $M_c$ and $\sigma_c$ are estimated by the method proposed by Wang et al. (2025) and Li et al., (2026). The earthquake magnitudes shown in the magnitude-time plot have been perturbed by a uniformly distributed random error within the range [-0.05, 0.05].

**Figure 3**. Temporal evolution of the calibrated parameters in the ETAS model with spatially variable background seismicity ($\mathrm{ETAS}_{\mu(x,y)}$) and in the ETAS model with a constant background rate ($\mathrm{ETAS}_{\mu}$). For model calibration, the primary catalog spans from January 1, 1985, to the day before each monthly forecast window, and is preceded by a 15-year auxiliary catalog (from January 1, 1970, to December 31, 1984) to prevent misclassification of early events in the primary catalog as background events. Note that the parameters $D$ and $Q$ are specific to $\mathrm{ETAS}_{\mu(x,y)}$ and are not included in $\mathrm{ETAS}_{\mu}$.

**Figure 4**. Time series of cumulative information gain (CIG) of $\mathrm{ETAS}_{\mu(x,y)}$ over $\mathrm{ETAS}_{\mu}$ in pseudo-prospective forecasting experiments at the CSES from 2000 to 2023. Results are shown for six spatial resolutions, five magnitude thresholds ($M_t$), and three strategies for assigning $P(n_i)$ when the observed value $n_i$ in the $i$-th triangle exceeds the forecasted maximum. The green,

black, and orange curves correspond to strategies (1), (2), and (3) described in Figure 1, respectively. The pseudo-prospective forecasting experiments use calendar months as time windows, resulting in a total of 288 time windows over the study period. The horizontal black dashed line indicates zero cumulative information gain.

**Figure 5**. Mean information gain (MIG) of $ETAS_{\mu(x, y)}$ over $ETAS_{\mu}$ across a total of 288 time windows during the study period in pseudo-prospective forecasting experiments for the CSES from 2000 to 2023, at a spatial resolution of 195 $km^2$. The information gain used to compute the mean information gain is based on the power law extrapolation method. Red colors in the color bar indicate positive mean information gain, meaning that $ETAS_{\mu(x, y)}$ outperforms $ETAS_{\mu}$.

**Figure 6**. Probability distributions of earthquake counts forecasted by (dashed line) $ETAS_{\mu}$ and (solid line) $ETAS_{\mu(x, y)}$ for May 2008 across the entire CSES region, under different magnitude thresholds ($M_t$). The $M_S$ 8.0 Wenchuan earthquake on May 12 and its aftershocks (marked by the red vertical line) generate an out-of-range situation as shown in Panel (b) of Figure 1, which corresponds to a single time-window slice taken from the last row of Figure 4. All plotting elements are consistent with those in Figure 1, except that the $ETAS_{\mu}$ data points are shown as solid gray triangles for visual distinction. The Data and Resources section provides links to access the forecasted probability density functions of both models for all forecast windows and magnitude thresholds across the full CSES region.

**Figure 7**. Frequency distribution function of earthquake count per grid cell over the ensemble of grid cells, aggregated over all 288 forecast windows for the China Seismic Experimental Site. Each point represents the number of grid cells (on the $y$-axis) that experienced a specific number of earthquakes (on the $x$-axis), for a specific spatial resolution and magnitude threshold. In other words, the curves show the empirical non-normalized probability density function of earthquake counts per cell. The black, orange, and red fitted lines denote the power law fits for the full study period, the pre-Wenchuan period, and the post-Wenchuan period, respectively. The corresponding exponents are denoted by $a_t$, $a_b$, and $a_a$. Consistent with Figure 1, the blue solid line represents the adaptive Gaussian kernel density estimate, while the dashed lines show the fitted power law function ($P(n) \sim n^{-(1+a)}$). For each combination of spatial resolution and magnitude threshold, the mean and standard deviation of the fitted exponent $a$ were estimated

from 200 bootstrap resamples.

**Figure 8**. Histograms of the 25 mean power law exponents *a* shown in Figure 7 for the total (top panel), pre- (middle panel) and post- (bottom panel) Wenchuan 2008 earthquake and their Gaussian fit.